\documentclass[journal=jacsat,manuscript=article, layout=twocolumn
            ]{achemso}

\usepackage{adjustbox}
\usepackage{amsmath}
\usepackage{amssymb}
\usepackage{bm}
\usepackage[labelformat=simple,labelsep=period,skip=3pt]{caption}
\usepackage{chemfig}
\usepackage{comment}
\usepackage{dcolumn}
\usepackage{float}
\usepackage{graphicx}
\usepackage[colorlinks, pdfpagelabels, pdfstartview=FitH, bookmarksopen=true, bookmarksnumbered=true, linkcolor=black, plainpages=false, citecolor=black, urlcolor=black]{hyperref}
\usepackage{lipsum}
\usepackage{makecell}
\usepackage[version=3]{mhchem} 
\usepackage{multirow}
\usepackage{nicefrac}
\usepackage{siunitx}
\usepackage{stmaryrd}
\usepackage{subcaption}
\usepackage{textcomp}
\usepackage[textsize=footnotesize]{todonotes}

\DeclareSIUnit{\atom}{atom}
\DeclareSIUnit{\angstrom}{\AA}

\author{Zsuzsanna Koczor-Benda}
\affiliation{Department of Chemistry, University of Warwick, Coventry, CV4 7AL, United Kingdom}

\author{Joe Gilkes}
\affiliation{Department of Chemistry, University of Warwick, Coventry, CV4 7AL, United Kingdom}
\alsoaffiliation{Centre for Doctoral Training in Modelling of Heterogeneous Systems, University of Warwick, Coventry, CV4 7AL, United Kingdom}

\author{Francesco Bartucca}
\affiliation{Department of Chemistry, University of Warwick, Coventry, CV4 7AL, United Kingdom}

\author{Abdulla Al-Fekaiki}
\affiliation{Department of Chemistry, University of Warwick, Coventry, CV4 7AL, United Kingdom}

\author{Reinhard J. Maurer}
\email{r.maurer@warwick.ac.uk}
\affiliation{Department of Chemistry, University of Warwick, Coventry, CV4 7AL, United Kingdom}
\alsoaffiliation{Department of Physics, University of Warwick, Coventry, CV4 7AL, United Kingdom}

\title{Structural bias in three-dimensional autoregressive generative machine learning of organic molecules}

\begin{document}

\begin{tocentry}
\includegraphics[width=3.25in]{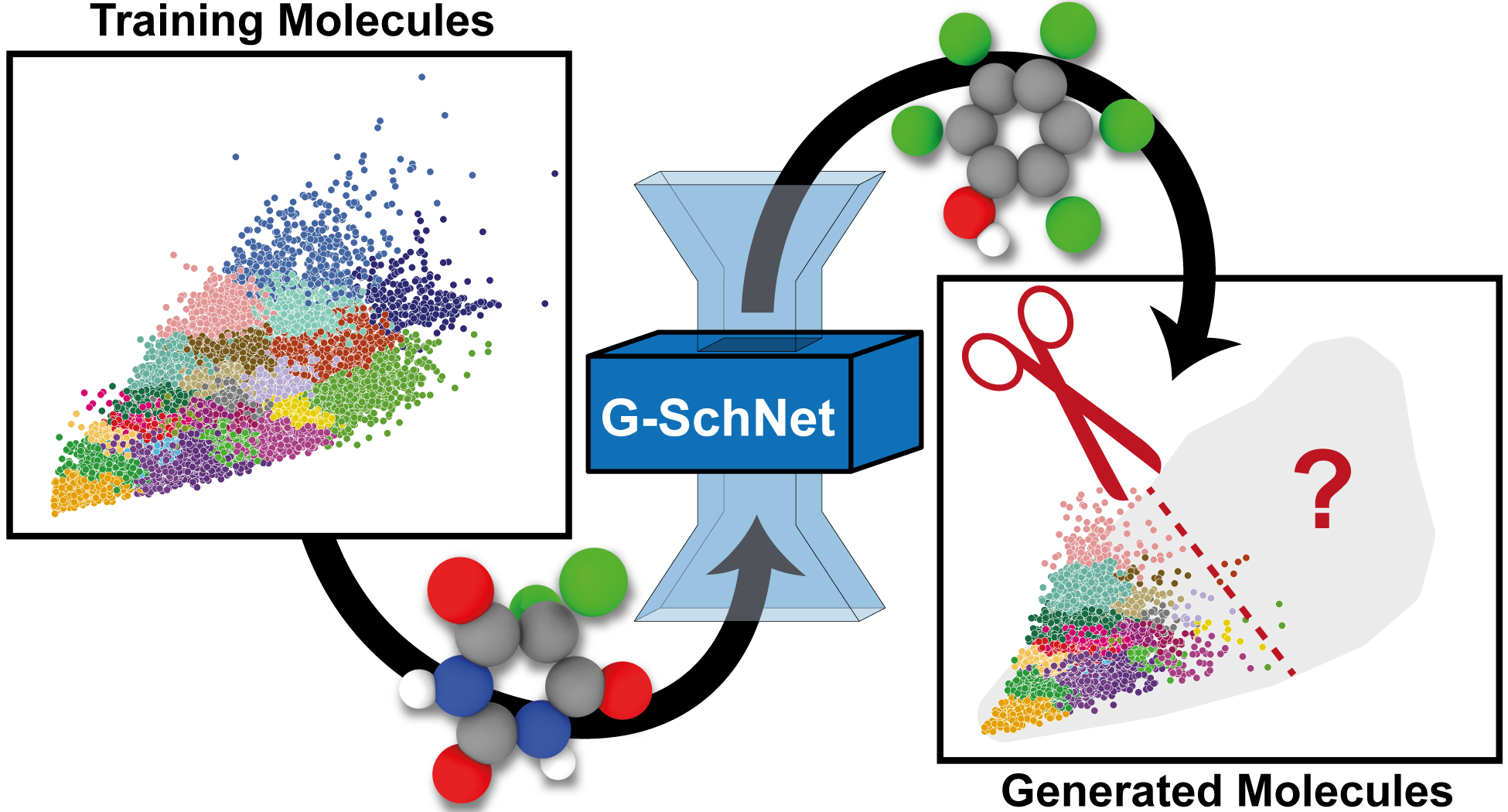}
\end{tocentry}

\begin{abstract}
A diverse range of generative machine learning models for the design of novel molecules and materials have been proposed in recent years. Models that are able to generate three-dimensional structures are particularly suitable for quantum chemistry workflows, enabling direct property prediction. The performance of generative models is typically assessed based on their ability to produce high rates of novel, valid, and unique molecules. However, equally important is the ability of generative models to learn the prevalence of functional groups and certain chemical moieties in the underlying training data, that is, to faithfully reproduce the chemical space spanned by the training data. 
Here we investigate the ability of the autoregressive generative machine learning model G-SchNet to reproduce the chemical space and property distributions of training datasets composed of large, functional organic molecules. We assess the elemental composition, size- and bond-length distributions, as well as the functional group and chemical space distribution of training and generated molecules.  By principal component analysis of the chemical space, we find that the model leads to a biased generation that is largely unaffected by the choice of hyperparameters or the training dataset distribution, producing molecules that are, on average, more unsaturated and contain more heteroatoms. Purely aliphatic molecules are mostly absent in the generation. We further investigate generation with functional group constraints and based on composite datasets, which can help partially remedy the model generation bias. Decision tree models can recognize the generation bias in the models and discriminate between training and generated data, revealing key chemical differences between the two sets. The chemical differences we find affect the distributions of electronic properties such as the HOMO-LUMO gap, which is a common target for functional molecule design. 

\end{abstract}


\section{Introduction}
Generative machine learning models (GMLMs) play an increasingly important role in chemistry. They support the exploration of the compositional and configurational space of molecules (chemical space) and the design of novel molecules and materials.~\cite{Elton2019, schwalbe-koda_generative_2020, bilodeau_generative_2022, anstine_generative_2023} 
Many variants of generative models for molecules exist based on a diverse range of architectures such as variational autoencoders (VAEs), generative adversarial networks, diffusion models, and autoregressive models. Typically, the performance of generative models is assessed based on their ability to generate novel, unique and valid molecules at a high rate, such that they can create diverse and realistic molecules. However, an equally important requirement is that the GMLM reproduces the chemical space and the structural distribution spanned by the training data. Good knowledge of the structural and functional group distribution that the GMLM produces, for example, is a prerequisite for inverse design and property-driven generative design applications, which typically aim to controllably find molecules beyond the original training data. In such applications, the original training data provides a starting point. This molecular distribution is then either biased \cite{gebauer2019symmetry, westermayr2023high} or constrained \cite{gebauer2022inverse, zhang_novo_2022} towards a specific region of chemical space to increase the likelihood of generating molecules that satisfy certain conditions or properties. 

Most GMLMs use text-based or image representations of molecules such as SMILES,~\cite{gomez-bombarelli_automatic_2018, putin_reinforced_2018} fingerprints,~\cite{kadurin_drugan_2017, tazhigulov_molecular_2022} or images~\cite{wang_image-based_2025} as these are most suitable when combined with existing generative algorithms. However, many applications require molecular representations in three-dimensional (3D) Euclidean space that capture bond distances, angles, and molecular conformation. This is particularly so when studying optoelectronic and vibrational properties of molecules that are sensitive to molecular conformation or when integrating GMLMs with quantum chemistry methods where 3D Cartesian positions are a necessary input to calculations. While 3D structures can be reconstructed from generative design based on structural fingerprints~\cite{fung_atomic_2022}, this adds an additional source of uncertainty and error. 
Several models have so far been proposed that directly generate 3D point cloud representations of molecules~\cite{gebauer2019symmetry,hoogeboom_equivariant_2022, xu_geometric_2023, hua_mudiff_2024, cornet_equivariant_2024}. While models based on normalizing flows and diffusion currently top the leaderboards when it comes to their ability to generate unique, novel, and valid molecules, they also come with considerable downsides. Existing diffusion models can often generate disconnected fragments at a high rate for large molecules.~\cite{jo_score-based_2022, xu_geometric_2023} In addition, molecular generation comes with considerable computational effort as the diffusion model acts simultaneously on all input coordinates and a large number of steps is required to converge the diffusion process. 
Autoregressive generation algorithms, on the other hand, build molecular structures atom-by-atom based on conditional probabilities of finding a specific element at a certain position in the presence of previously placed atoms. G-SchNet captures the elemental composition and the chemical environment of the atoms using the SchNet descriptor embeddings.~\cite{Schutt2017, Schutt2018} G-SchNet has been successfully used for a variety of applications, particularly in the context of conditional and biased property-driven design.~\cite{gebauer2022inverse, westermayr2023high, koczor-benda_generative_2025} For these applications, it is crucial that the generative model accurately represents the underlying distribution provided by the training data.

The ability of G-SchNet to reproduce chemical features and functional groups of its training database has been analysed in previous publications \cite{gebauer2019symmetry,joshi20213d,gebauer2022inverse}. These studies were mostly focused on the QM9 \cite{RamakrishnanSD14} training dataset, which has a relatively restricted chemical space, containing elements H, C, N, O, F, and only up to 9 non-hydrogen atoms per molecule. Training sets containing more diverse and larger molecules have been used for scaffold-based design \cite{joshi20213d} and property-driven design \cite{westermayr2023high, koczor-benda_generative_2025}. 
Westermayr \textit{et al.} \cite{westermayr2023high} trained G-SchNet on the OE62 dataset of  61,489 crystal-forming organic molecules.~\cite{stuke_atomic_2020} The OE62 dataset features large chemical and structural diversity with molecules varying in size from 3 to over 150 atoms and containing up to 16 elements. Westermayr \textit{et al.}~\cite{westermayr2021physically}  reported differences in elemental composition and subtle deviations in the bond distance and angle distributions between the training data and the generated structures. However,  fundamental quasiparticle gaps predicted for the unbiased generated molecules differed from the original training dataset by several electronVolt. Nevertheless, the property-driven design towards lower fundamental gaps based on iterative fine-tuning of G-SchNet was successful, as the independent property predictor that was employed remained largely valid for all generated molecules.  Koczor-Benda \textit{et al.}~\cite{koczor-benda_generative_2025} recently employed G-SchNet for the targeted design of molecules with optimal vibrational properties to act as THz upconverters in plasmonic nanojunctions,~\cite{xomalis_detecting_2021, chen_continuous-wave_2021} requiring the presence of a gold-thiolate group and simultaneous high infrared and Raman activity in the THz region. Unbiased generated molecules exhibited significant chemical differences from the training dataset with the consequence that the employed predictor for THz activity was not transferable to the generated molecules without retraining.

Here, we perform an in-depth analysis of the ability of the G-SchNet GMLM to generate datasets of molecules with up to two hundred atoms with structural and chemical properties that are consistent with the underlying training data. We do this for G-SchNet trained on the QM9 dataset,~\cite{RamakrishnanSD14, RuddigkeitJCIM12} the OE62 dataset,~\cite{stuke_atomic_2020} and a dataset of thiol molecules.~\cite{koczor-benda_molecular_2021, koczor-benda_generative_2025} Largely independent of the hyperparameters, G-SchNet consistently generates more undersaturated molecules than the underlying datasets, which is reflected in the elemental and molecular weight distribution. As revealed by clustering analysis, the generated molecular distribution lacks specific molecular motifs, such as large aliphatic structures, which affects different molecular properties. We find that resampling the training data to provide a more equal representation of structural features and functional groups in the training data only partially remedies the observed discrepancy between the chemical space covered by the training data and the generated molecules. 
Based on this analysis, we study the role of functional group constraints and training dataset composition on the chemical space distribution of generated molecules and how the structural bias during molecule generation affects property distributions. Bias in molecular generation can be used to discriminate original from generated molecules. 

\section{Methods}

\textbf{Generative machine learning}.
We use the schnetpack-gschnet package \cite{schnetpack-gschnet} for all model training and molecule generation. In this work, we employ three databases: the QM9 database of 134k molecules with 9 non-hydrogen atoms,~\cite{RamakrishnanSD14,RuddigkeitJCIM12} the OE62 database of 61,489 crystal-forming organic molecules extracted from the Cambridge Crystal Structure Database,~\cite{stuke_atomic_2020} and a dataset of about 3,000 gold-thiolate molecules,~\cite{koczor-benda_molecular_2021, koczor-benda_generative_2025} henceforth referred to as the `THz database'.   All datasets are public and provide molecules in their relaxed, equilibrium geometry as predicted by density functional theory (DFT).
For QM9, we use 50,000 (50k) molecules for training, 5k for validation, and 76k for testing. We employ default hyperparameter settings of the G-SchNet QM9 experiment unless stated otherwise. We set the number of molecules to generate as 60k, and the maximum number of atoms per molecule to 35 (following Ref.~\citenum{gebauer2019symmetry}). For the learning curve, the same validation set (5k) and test set (76k) were used across all models.  
For the OE62 database, we use 45k molecules for training, 4.5k for validation, and 12k for testing. The placement cutoff and the covalent radius factor in G-SchNet are increased (to 2.6 \AA\ and 1.3, respectively), to accommodate the increased variety of chemical elements. As in Ref. \citenum{westermayr2023high}, we reduce the number of random atom placement trajectories sampled to 5, to reduce computational cost. We show later that this does not significantly affect the generation. The parameters used for generation are the same as for QM9, except for the maximum number of atoms that was set to 200, to account for the larger size of training molecules. For the constrained gold-thiolate molecule generation, we combine the OE62 dataset with the THz dataset using the same settings as for OE62-only training and generation. Further details on the hyperparameters of G-SchNet for training and generation are provided in the Supplemental Information (SI).

\textbf{Molecular analysis}.
Generated molecules are filtered using the approach from Ref. \citenum{westermayr2023high}, which removes duplicates and disconnected structures. Unless stated otherwise, only these filtered generated molecules are included in the molecular analysis. All molecules from the QM9 and OE62 training databases are included in the analysis. For elemental composition analysis, pairwise distance distribution analysis and generation of canonical SMILES~\cite{SMILES} representations, the Open Babel~\cite{Open_Babel} package was used. Ring types and counts, as well as functional groups were identified with the RDKit package \cite{RDKit}. We note that in many cases, the canonical SMILES strings did not convert to valid RDKit molecules, even for the training databases, therefore these molecules are not included in further analysis. Further information is provided in the SI. HOMO-LUMO gaps were predicted for the training and generated molecules using the SchNet+H model~\cite{westermayr2021physically}. 

\textbf{Clustering in latent chemical space}.
Visual representations of the chemical space spanned by molecules within training and generated databases are created through the application of principal component analysis (PCA) on two high-dimensional molecular descriptors, following the same formalism introduced in Ref. \citenum{westermayr2023high}. The first descriptor, referred to as the structural descriptor, is an averaged SOAP (smooth overlap of atomic positions)~\cite{SOAP} descriptor, obtained with the DScribe package.\cite{HIMANEN2020} This results in a 57,792-dimensional encoding of the average atomic environment around each atom for each molecule in a given database. The second descriptor, referred to as the bonding descriptor, is a bespoke descriptor formed from features relating to molecular bonding extracted by the Open Babel and RDKit software packages.\cite{Open_Babel, RDKit} 592 of these features make up the bonding descriptor, ranging from simple quantities such as the number of each element in a molecule, to more complex measures such as the number of aromatic rings of a given size in each molecule. 

After calculating these descriptors for every molecule in a database, independent PCAs are fit for each descriptor. In cases where multiple databases need to span the same latent chemical space for comparison, these are fit across the concatenation of these databases. Molecules are plotted in the latent space formed by the first principal component (PC) from each descriptor to evaluate the extent to which datasets align. We have carefully assessed that both descriptor spaces efficiently capture their respective covariances in a single PC, providing more than 66\% covariance in the first principal component. We apply a 2D kernel density estimate (KDE) to the molecules in these latent spaces to identify areas of localisation. To aid with the analysis of these densely-packed latent spaces which often contain tens of thousands of molecules, we cluster molecules within this latent space. Again, when comparison between datasets is required, this is performed over the concatenation of these datasets such that clustering applies across datasets, and molecules in a given cluster within one dataset are similar in structure and bonding to molecules in the same cluster in another dataset. As in Ref. \citenum{westermayr2023high}, we use a mixture of the BIRCH~\cite{BIRCH} and agglomerative clustering~\cite{agglomerative} algorithms, as implemented in the scikit-learn package~\cite{scikit-learn}, to allow for finding clusters with different sizes.

We later resample the OE62 training dataset within the latent space by dividing the space created by the maximum and minimum bounds of the first PCs from each descriptor into an evenly spaced grid, then randomly sampling a single molecule from each grid rectangle. The number of returned molecules is controlled by the resolution of this grid (how many divisions are created along each axis), $n_{\textrm{grid}}$, chosen by fitting a 2D KDE to the sampled molecules and optimising for a lack of spatial localisation.

\textbf{Decision tree discriminators}.
We train decision trees to discriminate between training and generated molecules. The same number of molecules were sampled randomly from training and generated databases, 20\% of molecules were used as test set, while the remaining molecules were split in 90\%/10\% for train and validation sets for finding the optimal depth of decision trees. As feature vectors, we tested Morgan fingerprint vectors with different radii and number of bits (see SI), 
generated with the RDKit package \cite{RDKit}.  We also tested adding total heavy atom counts and individual element counts to the feature vectors (see SI), which increases accuracy scores in the case of QM9, but has no notable effect in the case of OE62. Consequently, these were used only for QM9. The tree depth parameter values were optimised for each dataset (see SI) to achieve the highest accuracy score on the validation set. For testing different G-SchNet models trained on OE62 with different G-SchNet hyperparameter settings, we train decision trees using the same type of feature vector (Morgan fingerprints only) and tree depth parameter (9) that were found to be optimal for the optimised G-SchNet model used to generate molecules from OE62 (see SI). The scikit-learn package \cite{scikit-learn} was used for model training.

\section{Results and Discussion}

\subsection{Small molecule generation}

Molecules generated by G-SchNet trained on the QM9 database have already been analysed in the original G-SchNet publication.~\cite{gebauer2019symmetry} Here we only briefly review and expand that analysis to emphasise some key points and provide more detailed insights that will enable us to compare models trained on datasets of larger and more complex molecules in the following sections.

\begin{figure}
    \centering
    \includegraphics[width=0.8\linewidth]{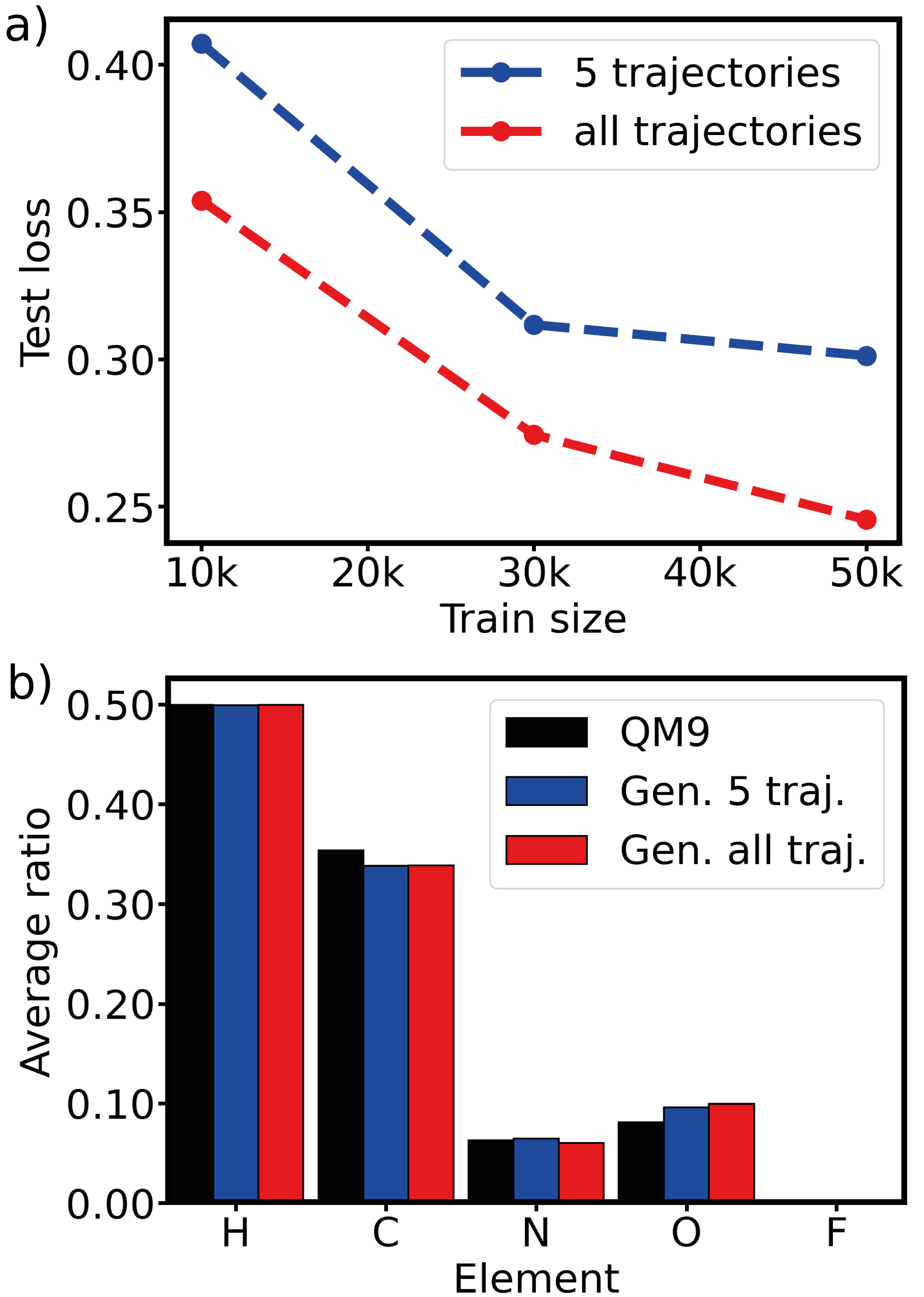}
    \caption{a) G-SchNet learning curve on the QM9 dataset, using 5 randomly sampled atom placement trajectories per molecule or all possible trajectories. b) Elemental composition of the generated molecules compared to QM9.}
    \label{fig:qm9_learning_curve}
\end{figure}

Trained on QM9, G-SchNet reproduces the atom number distribution per molecule well (Fig. S2), generating only slightly larger molecules than those present in QM9 (18.6 vs. 18.0 atoms on average). However, the molecular weight distributions reveal that a significant number of generated molecules have larger molecular weight (extra peak at 140 g/mol, see Fig. S2) than molecules in QM9. Resampling the generated dataset to match the atom number distribution of the original training dataset does not have a significant impact on the molecular weight distribution (Fig. S2), suggesting that the proportion of heavier elements is inherently larger in generated molecules. By analysing the elemental composition, we find this is mainly due to oxygen and nitrogen atoms being slightly over-expressed while hydrogens and carbons are slightly under-expressed (Fig.~\ref{fig:qm9_learning_curve}b and Fig. S3). This behaviour was already noted in the original G-SchNet paper.~\cite{gebauer2019symmetry}

\begin{figure}
    \centering
    \includegraphics[width=0.8\linewidth]{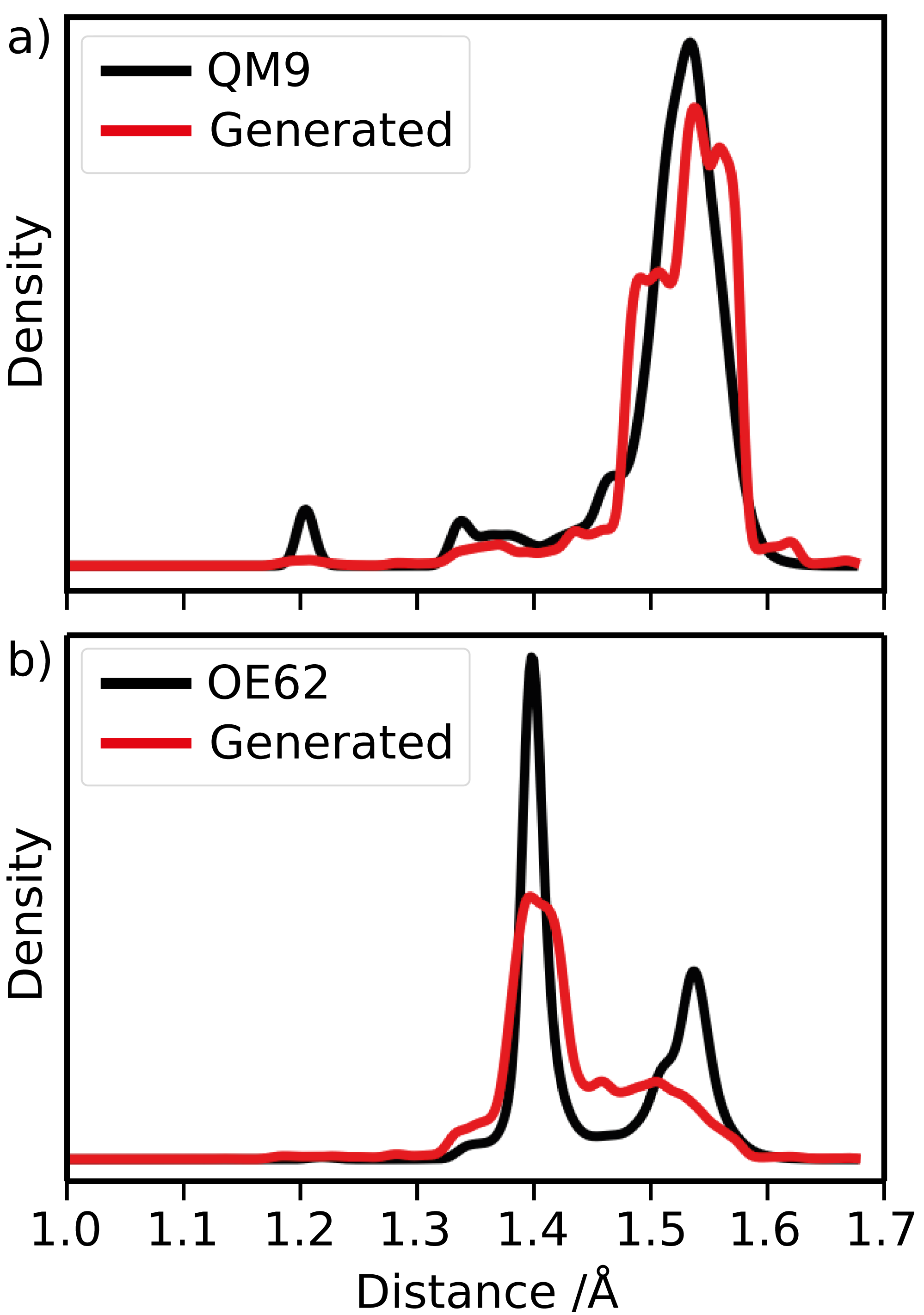}
    \caption{Distribution of bonded C-C distances for a) the QM9 and b) the OE62 training set and corresponding generated molecules. Generated molecules are sampled to match the atom number distribution of the respective training sets.}
    \label{fig:qm9_c-c}
\end{figure}

The radial distribution functions of C-C and C-O distances have also been reported in Ref. \cite{gebauer2019symmetry}. Here we want to point out that although the overall agreement is impressive for general atom-pair distances, some discrepancies become apparent (Fig.~\ref{fig:qm9_c-c}a) when looking at only bond distances. It becomes clear that the model cannot fully reproduce the distribution of single, double, and triple bonds. For example, short C-C distances, especially C-C triple bonds are underrepresented in the generated database. Functional group analysis shows that C-C triple bonds are present in 14\% of molecules in QM9, but only in 3\% of generated molecules. Similarly, C-C aromatic bonds are present in 13\% of QM9, but only in 2\% of generated molecules.

We trained decision tree models to classify molecules as training or generated based on molecular fingerprints and element counts (Fig. S5). Our best model achieves an accuracy score of 0.74 on previously unseen molecules. Among the most important features to classify molecules as either generated or belonging to the training data (top five most important features are shown in the SI, Fig. S6) are the number of non-hydrogen atoms, the number of nitrogen atoms, and the number of carbons involved in a triple bond. The first is easy to understand: since only the total number of atoms was restricted, molecules with more than 9 non-hydrogen atoms are present in the generated set; this clearly sets them apart from QM9, which contains at most 9 non-hydrogen atoms. Inaccuracies in elemental composition have also been discussed in the previous paragraphs. The presence of a triple bonded carbon in molecules with $\leq$ 9 heavy atoms likely results in a classification as training, in line with our observations that C-C triple bonds are underrepresented in generated molecules.

Average ring counts of 3-6 membered rings per molecule have been reported in Ref. \citenum{gebauer2019symmetry}. These show an increase in overall ring count, due to an increase in 3 and 4-membered rings. In Fig. S4, we additionally analyse the proportion of molecules having different numbers of aromatic, unsaturated aliphatic and saturated rings. Fewer of the generated molecules have aromatic and unsaturated aliphatic rings, while the presence of saturated rings increased noticeably, in particular, the proportion of molecules containing 2 or 3 saturated rings increased. 

For total dipole moment, isotropic polarisability, and electronic spatial extent, Gebauer \textit{et al.} \cite{gebauer2019symmetry} reported SchNet property predictions for training and generated molecules, while HOMO-LUMO gaps were calculated by DFT. The distributions for the generated molecules and QM9 are quite similar for all four properties, indicating that the minor chemical differences between generated and training molecules do not affect property distributions significantly.

Finally, we investigate how quickly G-SchNet learns from training data (3D molecular structures), by analysing models trained on 10k, 30k and 50k molecules from QM9.
As the number of possible atom placement trajectories scales quadratically with the number of atoms, to limit the cost of training down, we restrict the sampling trajectories during training to five. To test how limiting the number of trajectories sampled during training influences molecule generation, we performed an ablation study with QM9. We set the number of randomly sampled atom placement trajectories to five and investigated different numbers of training molecules. The resulting learning curve is shown in Fig.~\ref{fig:qm9_learning_curve}a. When limiting the number of trajectories, the validation and test losses are also calculated for the same number of random trajectories. While losses are increased with fewer trajectories, the elemental composition of the generated molecules is not significantly affected.

\subsection{Large and chemically diverse molecules}

As expected, the wider variety of elements and molecular sizes in the OE62 database presents a challenge for G-SchNet. Already the proportion of successful molecule generation to requested number of molecules is significantly lower for OE62 than for QM9. The details on the number of connected/unique/valid molecules, and how many can be converted to RDKit molecules, are given in Table S4.

\begin{figure*}
    \centering
    \includegraphics[width=\linewidth]{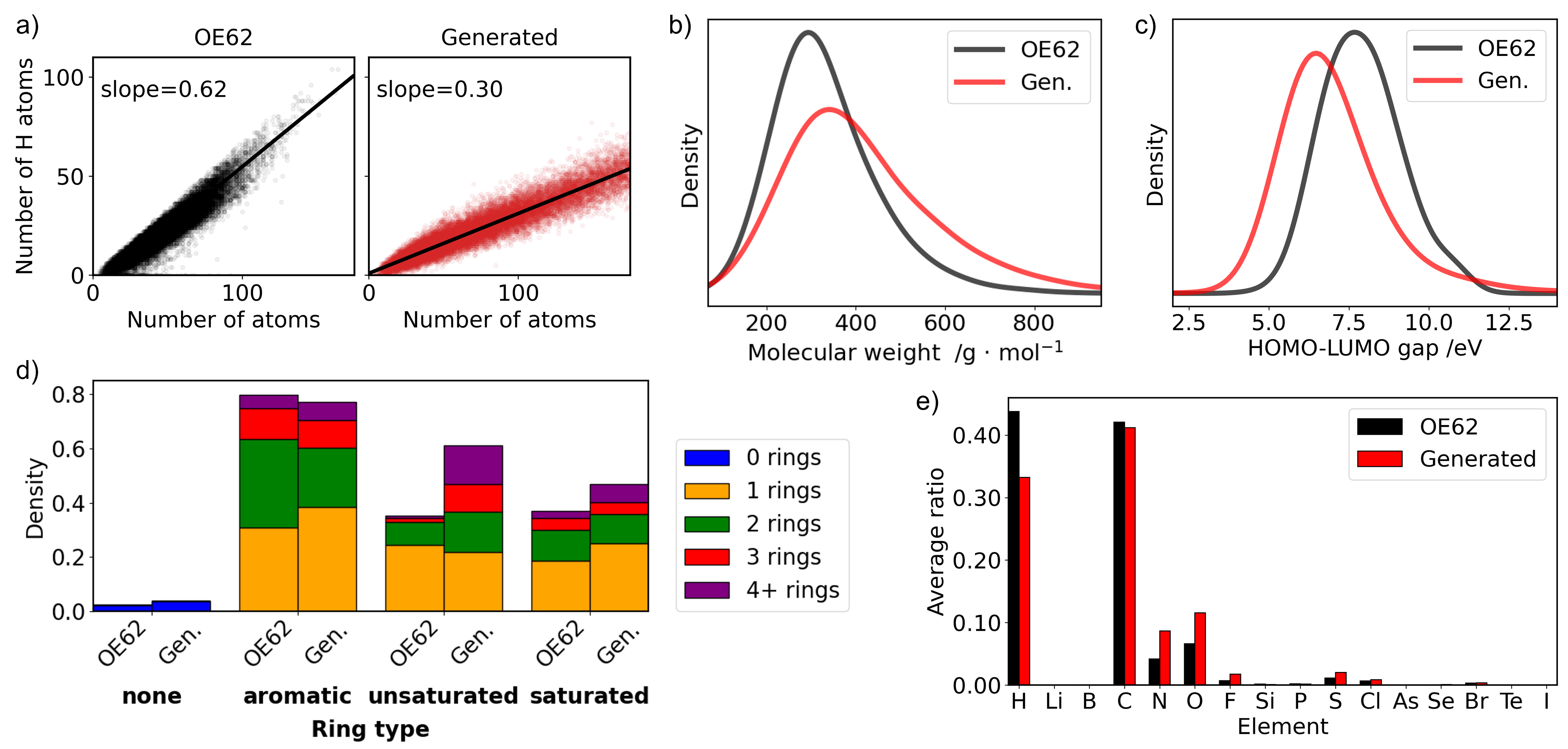}
    \caption{Analysis of the OE62 training database and corresponding generated molecules: a) Hydrogen atom count vs. total atom count, b) Molecular weight distributions, c) HOMO-LUMO gap distributions as predicted by the SchNet+H \cite{westermayr2021physically} model, d) Type and number of rings in molecules, e) Elemental composition. In b)-e) the generated database is sampled to match the atom count distribution of the training set.}
    \label{fig:oe62_natoms}
\end{figure*}

The size distribution of generated molecules is significantly different from that of the training database, in contrast to the QM9 case. The generated molecules are typically much larger than the molecules contained in the training dataset and also contain a smaller proportion of hydrogen atoms (Fig. \ref{fig:oe62_natoms}a and Fig. S8), indicating that heavier atoms and unsaturated bonds are over-expressed. To match the size distribution of the training set as closely as possible, it is necessary to either limit the maximum number of atoms to a significantly smaller number than the maximum number of atoms in OE62 or to resample the generated molecules according to the size distribution of the training database. Fig. \ref{fig:oe62_natoms}b shows that even after resampling according to the size distribution, the molecular weight distribution of the generated set is broader and shifted to larger values than that of OE62. This suggests that the generated molecules are more likely to contain heavier atoms, which is in line with observations  for QM9.

The elemental composition (Fig. \ref{fig:oe62_natoms}e) reveals that N, O, F, S, and Cl are all over-expressed. Interestingly, in contrast to our observations for the QM9 database, in this case, the longer bond distances corresponding to single C-C bonds (Fig. S8) 
are underrepresented compared to unsaturated bonds. This is also reflected in the increased proportion of molecules that contain multiple unsaturated aliphatic rings compared to OE62 (Fig. \ref{fig:oe62_natoms}d). 

To check if these effects are inherent to G-SchNet or are an artefact of filtering the generated molecules for validity and uniqueness, we performed the same analysis for raw generated molecules (Figs. S7-S9). Filtering has a negligible effect across the investigated distribution and does not affect any of the above-mentioned trends.

\begin{figure*}[t]
    \centering
    \includegraphics[width=\linewidth]{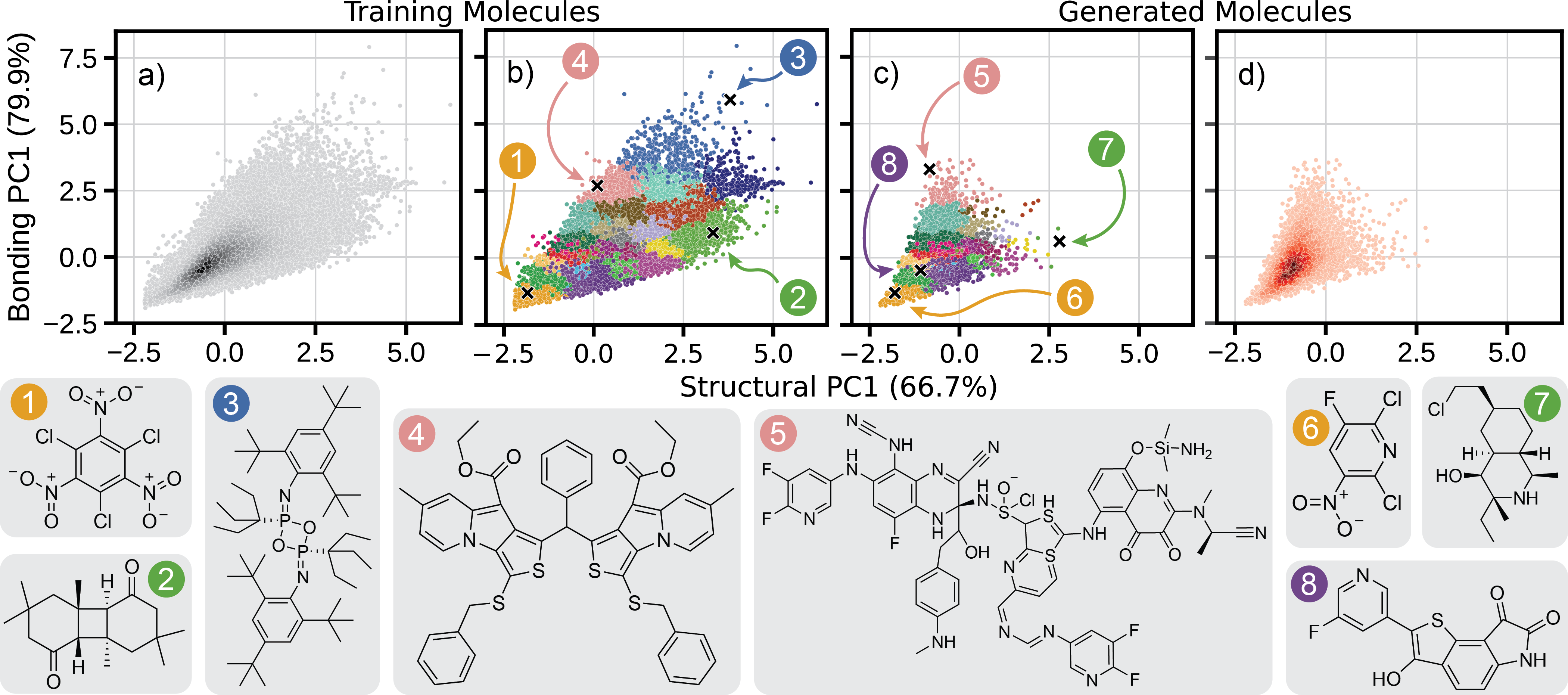}
    \caption{Latent chemical space covered by the molecules of the OE62 dataset and the generated molecules (sampled according to OE62’s atom count distribution), formed by PCA on descriptors encoding chemical structure and bonding. Percentages on axis labels indicate the proportion of the overall variance in each descriptor captured by the respective PC. a) and d) are KDE heatmaps of the training and generated molecules respectively, where darker areas correspond to a higher population of molecules. b) and c) represent the same molecules, divided into clusters representing similar areas of the latent space. Eight molecules, spread across the training and generated datasets, are shown to highlight similarities and differences in regions of the space covered by the datasets.}
    \label{fig:oe62_pca_heatmap}
\end{figure*}

The latent chemical space occupied by the molecules of the training and generated datasets is shown in Fig. \ref{fig:oe62_pca_heatmap}. Within this latent space, the generated molecules mimic the majority of the training set molecules, following the edges of the space outlined by OE62 and capturing the area of highest density (Fig. \ref{fig:oe62_pca_heatmap}a and \ref{fig:oe62_pca_heatmap}d). However, there are some differences between the two distributions; while the area between $\left( -2.5, 1.5 \right)$ in both PCs is replicated well, the generated molecules do not extend into the same high-value areas of either PC that are achieved in the training dataset. This is somewhat sensible, as the well-replicated regions lie close to the high-density region of the latent space shown in Fig. \ref{fig:oe62_pca_heatmap}a, while the unreplicated regions lie in spaces of comparatively low density. We verified that this lack of coverage is not caused by the resampling of generated molecules based on the size distribution of the training database by creating a similar latent space map of the unsampled generated molecules (Fig.~S10).

Clustering analysis on these latent spaces was also carried out, as shown in Fig. \ref{fig:oe62_pca_heatmap}b and \ref{fig:oe62_pca_heatmap}c. A selection of molecules from some of these clusters are also shown to highlight similarities and differences between regions of the latent spaces.

Molecules 1) and 6), from the training and generated datasets, respectively, show that the clustering approach successfully identifies similar clusters across both datasets; by probing the latent space at the same location in either dataset, structurally and chemically similar molecules can be extracted. This cluster in either dataset was characterised by small molecules with a high proportion of desaturation and aromaticity, as well as a significant proportion of heteroatoms per molecule. This is contrasted in molecules 2) and 7), which come from a cluster that consists of molecules with aliphatic rings, high saturation and few heteroatoms. G-SchNet found the generation of these types of molecules particularly difficult, echoing our earlier observations regarding the increased prevalence of heteroatoms and unsaturated bonds across the generated dataset.

Clusters around molecule 3) were not represented at all in the generated dataset; this is likely due to molecules in this area of the latent space possessing repeated structural motifs and functional groups, in addition to lines of symmetry. These molecules are usually synthesised through repeated applications of the same processes, making them difficult to create with G-SchNet as it generates molecules stochastically and is unlikely to repeat the same generative path multiple times per molecule. The vast majority of generated molecules from the high-density region shown in Fig. \ref{fig:oe62_pca_heatmap}d were instead similar to molecule 8), containing a handful of aromatic or otherwise unsaturated rings and 4-8 heteroatoms --- usually N, O, F, S, and Cl.

Within the replicated area of the latent space, the overlap is not perfect however; some generated molecules at values of Bonding PC1 between $1.0-3.75$ exist at lower values of Structural PC1 than in the training dataset, giving rise to an area of the latent space that is covered in Fig.~\ref{fig:oe62_pca_heatmap}c but not in Fig.~\ref{fig:oe62_pca_heatmap}b. Molecule 5) is from this extended region, while molecule 4) is one of the closest molecules to this area in the training dataset. The clusters from which these molecules originate generally encompass a variety of large, extended molecules with a high degree of whole-molecule conjugation, but G-SchNet's tendency to over-express heteroatoms leads to highly substituted molecules such as 5), which are not present in the training database due to the complexity of such molecules. The entries in OE62 are real, crystal-forming molecules, while the molecules from this region of the generated dataset are so complex that they are unlikely to be formed either as natural products or through manual synthesis. There is potential to filter out these molecules by sampling generated databases with respect to the molecular weight distribution of the training dataset, rather than its size distribution, as shown in Fig.~S11.

We have also trained a decision tree discriminator for OE62, which has a much higher accuracy score than the one trained for QM9. (Table \ref{tab:losses}). The most important features that discriminate generated molecules from the OE62 dataset are mainly carbon atoms involved in different, e.g. unsaturated, aromatic or aliphatic bonds (see Fig. S15 for a depiction of the five most important Morgan fingerprint bits), which is in line with our observations that C-C bond distance distributions are not captured well by the model. The hydroxyl group is also among the most important features, and our analysis shows that this group is significantly enriched in generated molecules (present in 67\% of generated vs. 22\% of OE62 valid RDKit molecules).

In contrast to QM9, when trained on OE62, G-SchNet produces molecules with significantly different property distributions. The HOMO-LUMO gap distribution of generated molecules (Fig.~\ref{fig:oe62_natoms}c) is shifted to smaller values by about 2 eV  (see also Ref.~\citenum{westermayr2023high} for electron affinities and HOMO-LUMO gaps), which is explained by the lower level of saturation for generated molecules.

To see if the model can be improved, we tested different parameter settings for training G-SchNet as well as for generating molecules with the trained models (Table~\ref{tab:losses}). This included changing the loss function from the one available in G-SchNet that is a sum of element type loss and distance loss with 1:1 weight ratio by default.

\begin{table*}[]
    \centering
    \begin{tabular}{c|cccc}
       & Distance loss & Element type loss & Combined loss & Decision tree score\\
       \hline
        Original  & 0.234 & 0.161 & 0.394 & 0.892\\
        20 \AA\ cutoffs  & 0.234 & 0.166 & 0.400 & 0.900 \\
        10 trajectories  & 0.216 & 0.152 & 0.367 & 0.908\\
        1:3 loss ratio  & 0.271 & 0.162 & 0.189 & 0.911 \\
    \end{tabular}
    \caption{Test losses and accuracy scores of G-SchNet models trained with different parameter settings. The original model uses 10 \AA\ model and prediction cutoffs, 5 randomly sampled trajectories, and a distance: element type loss ratio of 1:1.}
    \label{tab:losses}
\end{table*}
From these tests, only increasing the number of randomly sampled trajectories resulted in improvement in the test losses (both in distance and element type loss). Increasing the model and prediction cutoff parameters does not affect the results, while changing the loss ratio to 1:3, thus giving more weight to the element type loss, only results in an increase of distance loss on the test set with no improvement in element type loss.

How these parameters affect the molecular weight distribution, elemental composition and bond distance distribution of generated molecules is reported in Figs.~S12-S13. In general, there are minor differences in elemental composition between the molecules generated with different models when compared to the discrepancy with the OE62 training set. 
The C-C distance distributions are practically the same with the different models, irrespective of considering all C-C distances or only bond distances, suggesting that this variation in distance loss is insignificant. The balanced accuracy scores of decision trees trained to discriminate each generated database from the OE62 database are also very similar (Table~\ref{tab:losses}). These findings show that tweaking the hyperparameters of the G-SchNet model cannot correct its inability to fully represent the chemical distribution of its training database.

Other sources of error are also possible; the SchNet descriptor, used internally within G-SchNet to represent local atomic environments with rotational, translational and permutational invariance,\cite{gebauer2019symmetry} may be at fault. Stark \textit{et al.} previously showed that even after iterative refinement of a training set, SchNet models using this descriptor failed to produce a smooth potential energy surface (PES) for the dissociative adsorption of \ce{H2} on a Cu(111) surface.~\cite{WojciechSchNet} In contrast, the equivariant PaiNN model\cite{Schutt2021} yielded a smooth PES with the same data. While the molecules generated here are chemically very different to the tested systems, the inability of the SchNet descriptor to provide smooth representations with atomic coordinates may affect the ability of G-SchNet to recognise similar atomic groups when constructing molecules.

\subsection{\label{sec:flattening}Role of the training data distribution}

Structural and chemical discrepancies between the training data and the generated molecules may also arise by providing a biased and imbalanced training data set. Fig.~\ref{fig:oe62_pca_heatmap}a shows many of the training molecules in OE62 localise around a relatively small area of the latent space created by PCA. This localisation makes up around 75~\% of the training data, such that when G-SchNet learns conditional probability distributions for the next atom type, distance and position, they are inherently biased towards emulating the atoms in these molecules. This explains why the generated molecules in Fig.~\ref{fig:oe62_pca_heatmap}d are similarly localised. Molecules in the unexplored areas of the latent space are likely not present because their generation would require multiple improbable samplings of these distributions.

This effect may be mitigated by `flattening' the training distribution, such that the model is trained on a more balanced distribution of molecules (as defined by the latent space). This may increase the relative likelihood of generating molecules in underrepresented regions of chemical space. OE62 was therefore flattened by sampling the principal component subspace on a regular grid with $n_{\textrm{grid}}=350$ grid points to yield a new training database of 15,552 molecules. A new G-SchNet model was trained on this flattened database, and was used to generate 19,976 new molecules (5,112 after sampling according to the size distribution of the new training database).

\begin{figure*}[t!]
    \centering
    \includegraphics[width=\linewidth]{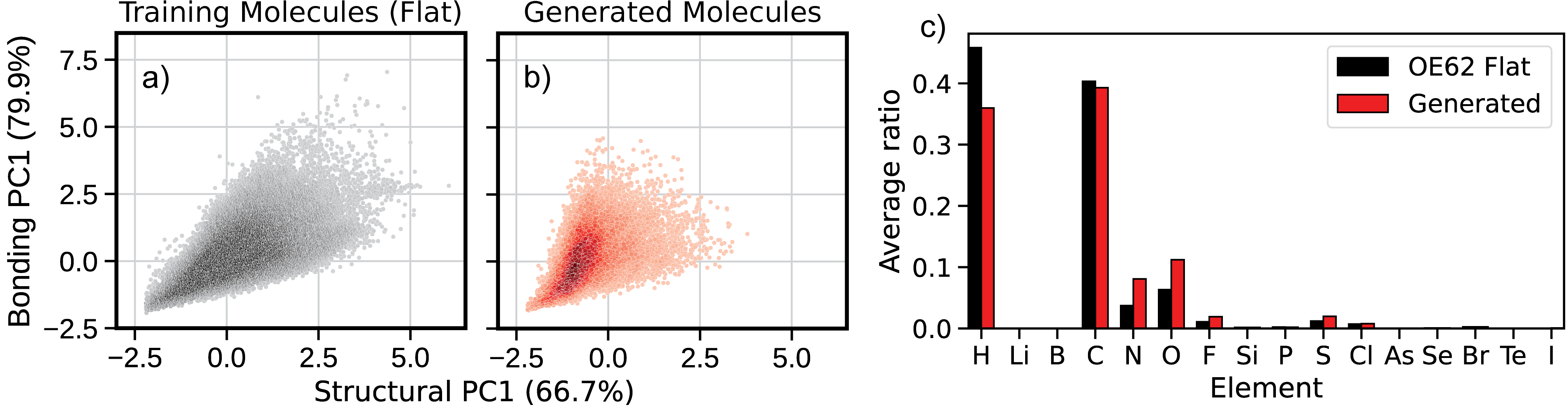}
    \caption{Analysis of the spatially flattened OE62 training database and corresponding generated molecules. Panels a) and b) show plots of the latent chemical space occupied by the databases, using the same PCA fits as in Fig.~\ref{fig:oe62_pca_heatmap}. Panel c) shows the difference in elemental composition between the two databases.}
    \label{fig:oe62_flat_pca}
\end{figure*}

KDE plots for the flattened training and generated molecules are shown in Figs.~\ref{fig:oe62_flat_pca}a and \ref{fig:oe62_flat_pca}b, alongside the elemental distributions for these databases in Fig.~\ref{fig:oe62_flat_pca}c. Comparing these spatial distributions of molecules to those in Fig.~\ref{fig:oe62_pca_heatmap}, the training and generated distributions cover the PCA space more equally. The high-density region is larger, and more molecules are generated with Structural PC1 $> 1$. As shown previously, this area of the latent space corresponds mostly to aliphatic molecules. Comparing the elemental distributions to those in Fig.~\ref{fig:oe62_natoms} reveals near-identical trends between the training and generated databases. Despite the flattened training database exhibiting a greater ratio of hydrogen atoms per molecule than the full OE62 database, the generated molecules still exhibit a significantly different hydrogen-to-carbon ratio.

Overall, while providing a dataset that is more evenly distributed in terms of the structural space and the bonding environments in molecules does slightly reduce the deviations in the structures of training and generated molecules, for the most part, large deviations remain. This affects the ability to sample novel molecules from the same space as the training data and will likely affect the ability to recover the associated distribution of molecular properties.

\subsection{Generation with functional group constraints}

Often, generative molecular design is based on property targets, but also on constraints with respect to the type and frequency of functional groups present in candidate molecules. For example, Koczor-Benda \textit{ et al.} have generated molecules with optimal THz upconversion efficiency when placed inside plasmonic nanojunction devices. Thiol groups ensure that the molecules strongly bind within the nanojunction via a gold-thiolate bond, making the presence of such a functional group a design prerequisite. To reliably judge the THz upconversion efficiency of candidate molecules, their properties need to be predicted base d on quantum chemistry calculations in the presence of the gold-thiolate group. Scaffold-based constraints can be imposed with G-SchNet by providing a filter that places the thiol group at the starting token of the autoregressive structure generation. However, most molecular datasets would contain few thiol groups and therefore it is highly likely that datasets will not provide a robust representation of relevant chemical environments. We address this issue by combining two chemically different databases; a dataset of about 3000 molecules containing gold-thiolate bonds,\cite{koczor-benda_molecular_2021, koczor-benda_generative_2025}, which provides the required chemical functionality, and OE62,\cite{stuke_atomic_2020}, which covers a wider chemical space and other potentially desirable chemical properties. We will call this combined dataset OE62+THz.

\begin{figure}
    \centering
    \includegraphics[width=\linewidth]{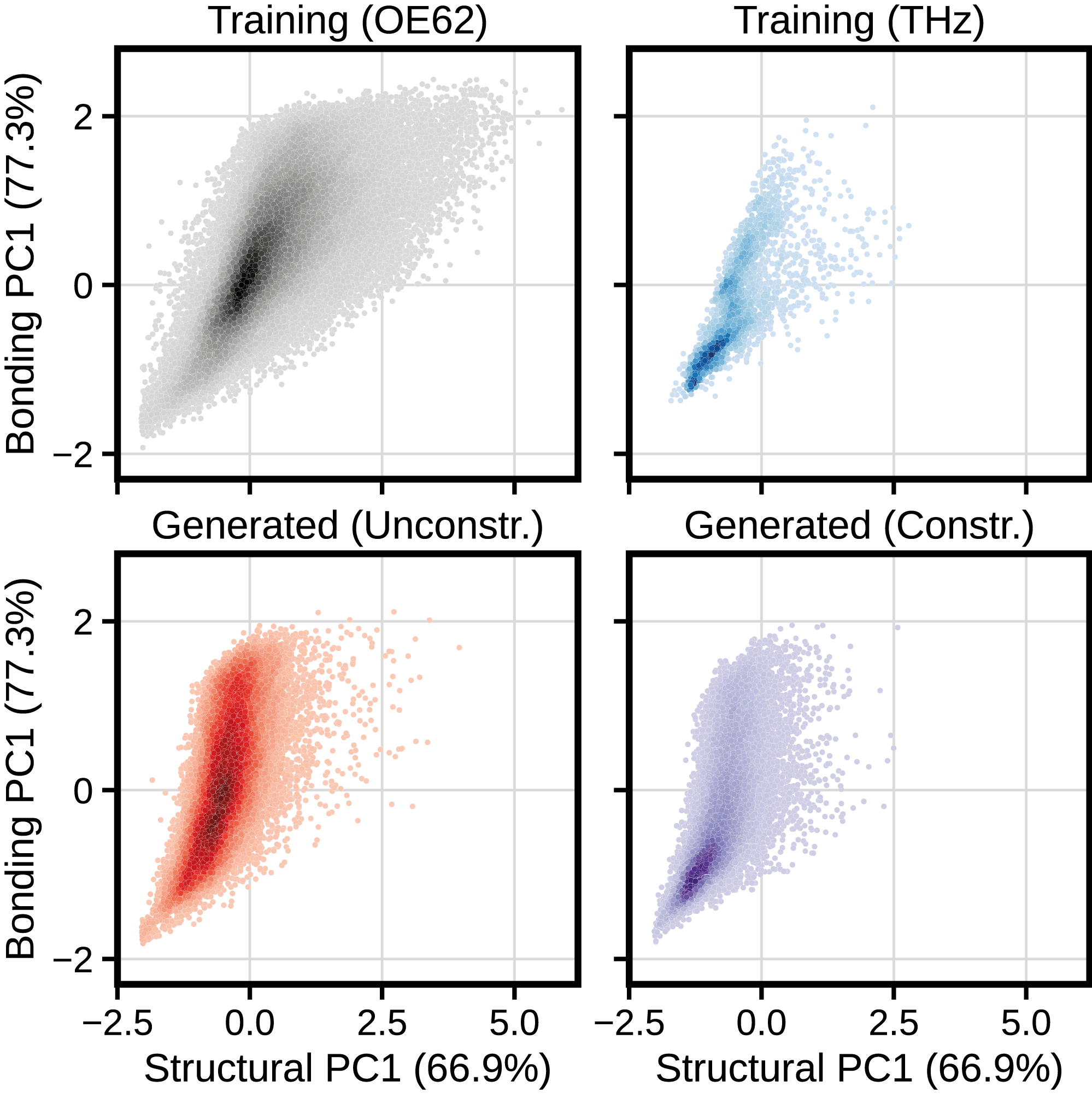}
    \caption{Density plots of latent chemical space covered by the molecules in the OE62 and THz training databases and the unconstrained and constrained generated databases, formed by PCA on descriptors encoding chemical structure and bonding. Darker colours indicate areas of higher density, as predicted by KDE. Percentages on axis labels indicate the proportion of the overall variance in each descriptor captured by the respective PC.}
    \label{fig:pca_constrained}
\end{figure}

When training an unconstrained G-SchNet model on the OE62+THz dataset (where the THz dataset contributes only about 5 percent of the training molecules), we observe similar trends in the data distributions as for OE62, i.e. a smaller proportion of H and C and an increased proportion of heteroatoms when compared to the training dataset. The latent chemical space covered by the molecules in these datasets (Fig. \ref{fig:pca_constrained})  shows that the THz subset of the dataset covers a significantly smaller region of the latent chemical space than OE62, with high concentrations of molecules located in different areas of latent space. Molecule generation based on the OE62+THz trained model explores a much smaller region than covered by OE62, but extends significantly beyond the region covered by the THz data. The generated molecules explore areas of latent space not covered by either of the training datasets. We note that the latent space in Fig.~\ref{fig:pca_constrained} is not the same as was shown in the earlier figure, due to PCs now being constructed for the bonding and structural descriptors across the combined OE62+THz training set, the unconstrained generated dataset and the constrained generated dataset.

Without imposing the scaffold constraint, about 4\% of generated molecules are gold-thiolates, whereas with the constraint 100\% of molecules have a gold-thiolate group. This is not fully visible in the latent space maps, likely because a single thiol group does not contribute much to either of the high-dimensional bonding and structural descriptors, resulting in the region covered by the molecules of the THz dataset being entirely contained within that of the molecules of OE62. We note that, in principle, a generated molecule can contain more than one thiol group, but we have not observed the generation of such a molecule.

The constraint leads to the generation of smaller molecules that are closer to those of the THz dataset than those of OE62. This is reflected in the latent space plots, where, although the extent of the regions covered by the unconstrained and constrained generated datasets is very similar, the high-density regions are entirely different. While the majority of unconstrained molecules are generated within the same high-density region of chemical space as was emphasised in OE62, most constrained molecules reflect the high-density region from the THz dataset instead. We emphasise that the constraint does not prevent G-SchNet from learning from OE62, as the lower-density regions of the constrained dataset still explore the same area as was also explored in the unconstrained case.

\begin{figure*}
    \centering
    \includegraphics[width=\linewidth]{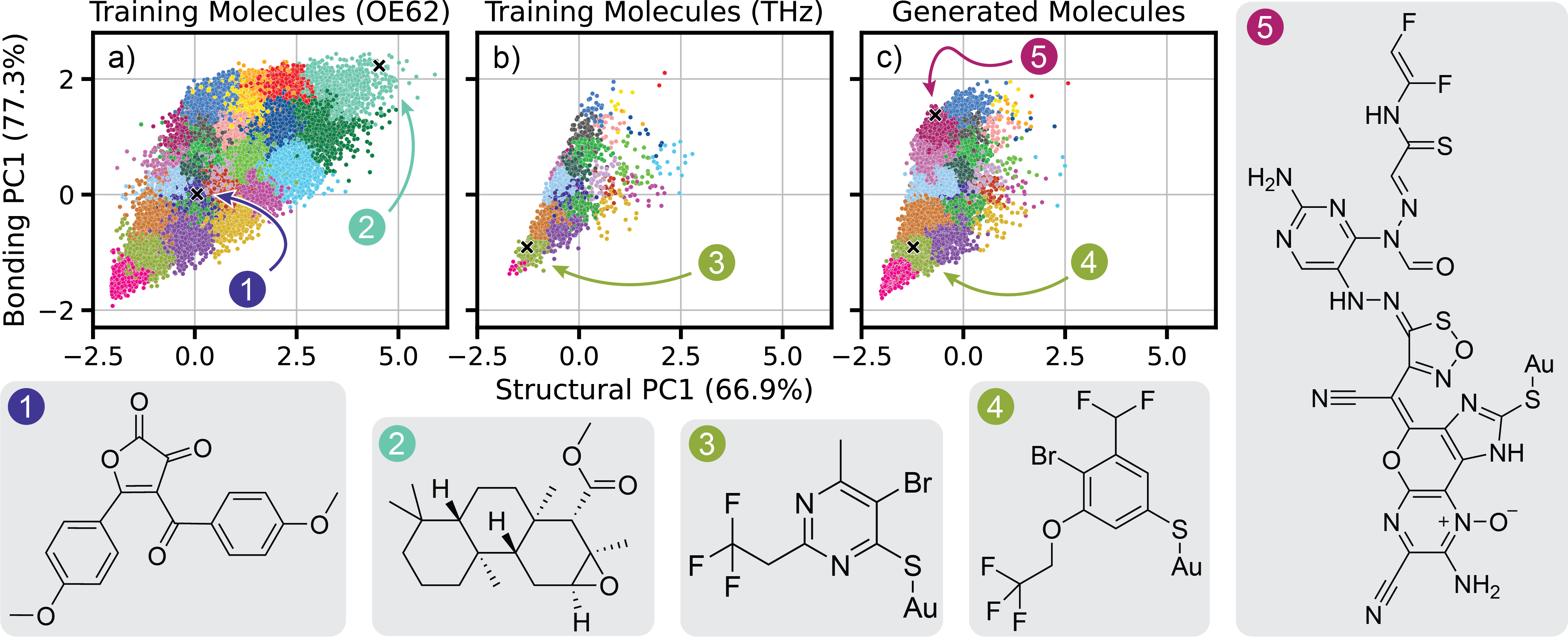}
    \caption{Clustering of molecules from a) the OE62 and b) the THz training databases, in addition to c) the constrained generated database, within the same latent chemical space formed by PCA as was shown in Figure \ref{fig:pca_constrained}. Examples of molecules from a subset of clusters are depicted.}
    \label{fig:clustering}
\end{figure*}

The same clustering analysis that was performed for the OE62-only trained model above was repeated with this new latent space to better identify trends in the generated molecules of the combined OE62+THz dataset, shown in Fig.~\ref{fig:clustering}. In this figure, molecule 1) represents part of the high-density region of OE62, which is characterised by multiple aromatic rings conjugated across the majority of each medium-sized (20-35 atoms) molecule, often with 3-7 heteroatoms. This cluster, and many surrounding it, are present in both the THz dataset and the constrained generated molecules, but its population is vastly reduced. Molecule 2), and others like it, are not present in the THz dataset or the constrained molecules; this is a highly aliphatic molecule with few heteroatoms though, and we previously established that G-SchNet struggles to generate such molecules. It is therefore not a surprise to see this region of the latent space sparsely covered in the generated dataset.

Molecules 3) and 4) represent characteristic examples from the high-density regions of the THz dataset and the constrained generated molecules, respectively. Both are smaller than molecule 1), reinforcing that the constrained model is, on average, generating smaller molecules. Of note is the high proportion of heteroatoms compared to carbon atoms in this area of the latent space, which turns out to be characteristic of the left edge (Structural PC1 $< 0.0$) of the space. This is echoed in molecules such as 5), which is much larger, but also contains many more heteroatoms. As in the earlier OE62-only model, molecules like 5) exist in an area of the latent space that was not covered by the molecules of the training set, as they are created by G-SchNet's tendency to over-express heteroatoms.

In summary, we find that by merging a database with a specific chemical functionality with a database that covers a wide area of chemical space, molecule generation with functional group constraints can be achieved for a larger area of chemical space, resulting in more diverse molecules. This is observed despite the remaining shortcomings of G-SchNet which lead to the insufficient coverage of chemical space by the generated molecules.

\section{Conclusion and Outlook}

We investigated the ability of the autoregressive three-dimensional molecular structure generation model G-SchNet to generate molecules that faithfully represent the structural distribution spanned by the training dataset. We provide a detailed analysis of the elemental distribution, the functional group composition and latent chemical space maps, which reveal that generated molecules are systematically biased compared to the training dataset. The bias leads to the overexpression of unsaturated carbon and heteroatoms compared to the training data and the suppression of aliphatic structural motifs. When visualised in latent chemical space maps, the molecules only cover a small region of chemical space compared to the training data. In addition, G-SchNet does not reliably represent the molecular size distribution of the training dataset, typically leading to the generation of many molecules larger than those in the training dataset, requiring additional subsampling of generated molecules with respect to the size or molecular weight distribution of the training dataset.

While structural bias effects in generated molecules are relatively subtle in small molecule datasets such as QM9, only minimally affecting the elemental distribution and the property distribution of the generated molecules, the effects are significant in larger molecules (up to 200 atoms) as represented in the OE62 dataset.~\cite{stuke_atomic_2020} Even without conditioning or fine-tuning, molecules generated by G-SchNet do not cover the same latent chemical space as the training dataset. Simple decision-tree-based classifier models can identify such bias and discriminate between training and generated molecules based on the presence of certain functional groups or the size of molecules. We introduce a strategy to ensure broader chemical space sampling despite the imposition of functional group constraints. Datasets with specific chemical functionalities that only cover a narrow region of chemical space can be complemented with diverse chemically unrelated datasets to create diverse molecules that reliably satisfy functional group constraints. By combining a small database of gold-thiolate-containing molecular structures optimised at DFT level (~3k molecules) with a comprehensive DFT database such as OE62, we demonstrate that G-SchNet is capable of learning features from both databases simultaneously while generating strictly thiol-containing molecules only.

While inaccuracies in reproducing size and elemental distributions of the training set might be negligible for some fields of applications, specific property prediction applications, such as electronic properties (e.g. HOMO-LUMO gaps), and property-driven design workflows require generative models to recover the training data distribution if a complementary molecular property prediction model is to be used to screen generated molecules. If this is not the case, property prediction models that may be trained and validated for the training dataset may become invalid for the generated molecules as they sit outside of the training data distribution. This can lead to uncontrolled errors in property prediction and results in a lack of control in property-driven workflows. This challenge has recently been discussed by Koczor-Benda \textit{et al.} in the context of property-driven design of molecules with optimal vibrational spectroscopic properties for THz detection~\cite{koczor-benda_generative_2025} 

In some cases, the tendency of G-SchNet to over-generate unsaturated and functionalised molecules might work in favour of generating molecules with desired properties even without conditioning or iterative biasing, such as for molecules with small HOMO-LUMO gaps.~\cite{westermayr2023high} In other cases, this may disqualify G-SchNet entirely for the task. While we here cannot provide an unambiguous conclusion of the causes of the inherent generation bias in the autoregressive model G-SchNet, we can confirm that an imbalance in the structural space spanned by the training dataset is likely only a small contributing factor. Insufficient expressivity of the descriptor will likely be a key factor that may be addressed in the future with the introduction of equivariant or vector-based learnable atom embeddings and latent feature representations in generative models. 

A critical question that remains is whether the bias we have identified in molecular generation with G-SchNet is also present in other state-of-the-art generative models based on other architectures, such as VAEs and normalizing flow diffusion. The current literature standard is to analyse molecule generation mostly based on molecular uniqueness, novelty, and validity, which may not be sufficient to fully assess generative design algorithms for their suitability to be employed in property-driven design workflows.

\section*{Data Availability}

Custom transform and generation functions to impose functional-group constraints in G-SchNet will be made available upon publication.
Code for the extraction of bonding features from molecular databases and obtaining the principal components of the structural/bonding descriptors has been released in our GSchNetTools package, available at \url{https://github.com/maurergroup/GSchNetTools}. Training and generated databases of molecules, along with principal components in both descriptor spaces, are available on figshare at \url{https://doi.org/10.6084/m9.figshare.28661504}.

\section*{Acknowledgments}

The authors acknowledge funding through the UKRI Future Leaders Fellowship programme (MR/X023109/1), a UKRI frontier research grant (EP/X014088/1), and the EPSRC Centre for Doctoral Training in Modelling of Heterogeneous Systems (EP/S022848/1) at the University of Warwick. High-performance computing resources were provided via the Scientific Computing Research Technology Platform of the University of Warwick, the EPSRC-funded HPC Midlands+ computing centre for access to Sulis (EP/P020232/1), and the Northern Ireland High Performance Computing (NI-HPC) service for access to Kelvin2 (EP/T022175/1).


\providecommand{\latin}[1]{#1}
\makeatletter
\providecommand{\doi}
  {\begingroup\let\do\@makeother\dospecials
  \catcode`\{=1 \catcode`\}=2 \doi@aux}
\providecommand{\doi@aux}[1]{\endgroup\texttt{#1}}
\makeatother
\providecommand*\mcitethebibliography{\thebibliography}
\csname @ifundefined\endcsname{endmcitethebibliography}
  {\let\endmcitethebibliography\endthebibliography}{}

\end{document}


\section{G-SchNet hyperparameter settings}

We have used G-SchNet version 1.0.0, commit: ecc48bcea85c88a415c6a98ee7f62db73ecc6031 (Nov 7, 2023). 
Tables S1 and S2 list some of the settings used for training and generation for the QM9 and OE62 datasets. All other settings were kept at the default values.

\begin{table}
    \centering
    \begin{tabular}{c|cc} 
      & QM9 & OE62   \\ \hline
      model\_cutoff   & 10.0  & 10.0  \\
      prediction\_cutoff   & 10.0 & 10.0  \\
      placement\_cutoff   & 1.7 & 2.6  \\
      use\_covalent\_radii   & true & true  \\
      covalent\_radius\_factor   & 1.1 &  1.3 \\
      lr   & 0.0001 & 0.0001 \\
      draw\_random\_samples   & 0 (generate all paths) & 5  \\
      batch\_size   & 5 &  5 \\
      num\_train   & 50,000 & 45,000  \\
      num\_val   & 5,000 & 4,500  \\ 
    \end{tabular}
    \caption{Training parameters used for models trained on QM9 and OE62 datasets. Units are the default units as expected by G-SchNet, i.e. \AA{} for distances.}
    \label{tab:training_params}
\end{table}

\begin{table}
    \centering
    \begin{tabular}{c|cc} \hline
      & QM9 & OE62   \\ \hline
    n\_molecules   & 60,000  & 60,000  \\
    batch\_size   & 1 &  1 \\
    max\_n\_atoms & 35 & 200  \\
    grid\_distance\_min & 0.7 & 0.7  \\
    grid\_spacing & 0.05 & 0.05  \\
    temperature\_term & 0.1 & 0.1  \\
    grid\_batch\_size & 0  & 0  \\ 
    \end{tabular}
    \caption{Generation parameters used for models trained on QM9 and OE62 datasets. Units are the default units as expected by G-SchNet, i.e. \AA{} for distances.}
    \label{tab:training_params}
\end{table}

\clearpage
\section{Additional data for model trained on QM9 training data}

\subsection{Generated Molecule Analysis}

\begin{table*}[]
    \centering
    \begin{tabular}{cc|ccccc}
       Trajectories & Train size & Target & Generated & Duplicate & Disconnected & Filtered\\
       \hline
        All & 10000 &  60000& 59832 & 0 & 8587 & 39256 \\
        All & 30000 & 60000& 58479 & 0 & 8609 & 38194\\
        All & 50000 & 60000 & 59165 & 0& 10271 & 37624 \\
        5 & 10000 &  60000& 59630 & 0 & 10575 & 38163 \\
        5 & 30000 & 60000& 59866 & 0 & 11537 & 38531\\
        5 & 50000 & 60000 & 59882 & 0& 10616 & 38190 \\
    \end{tabular}
    \caption{Number of unique and valid molecules generated with different G-SchNet settings, based on the QM9 training dataset.  }
    \label{tab:generated_stats}
\end{table*}

The QM9 training database contains 130831 molecules, out of which 128765 can be converted into valid RDKit molecules. The G-SchNet model trained on all atom placement trajectories and with a training data set size of 50000 generates 35551  molecules for which RDKit is able to generate valid SMILES strings. 

\begin{figure}
    \centering
    \includegraphics[width=0.7\linewidth]{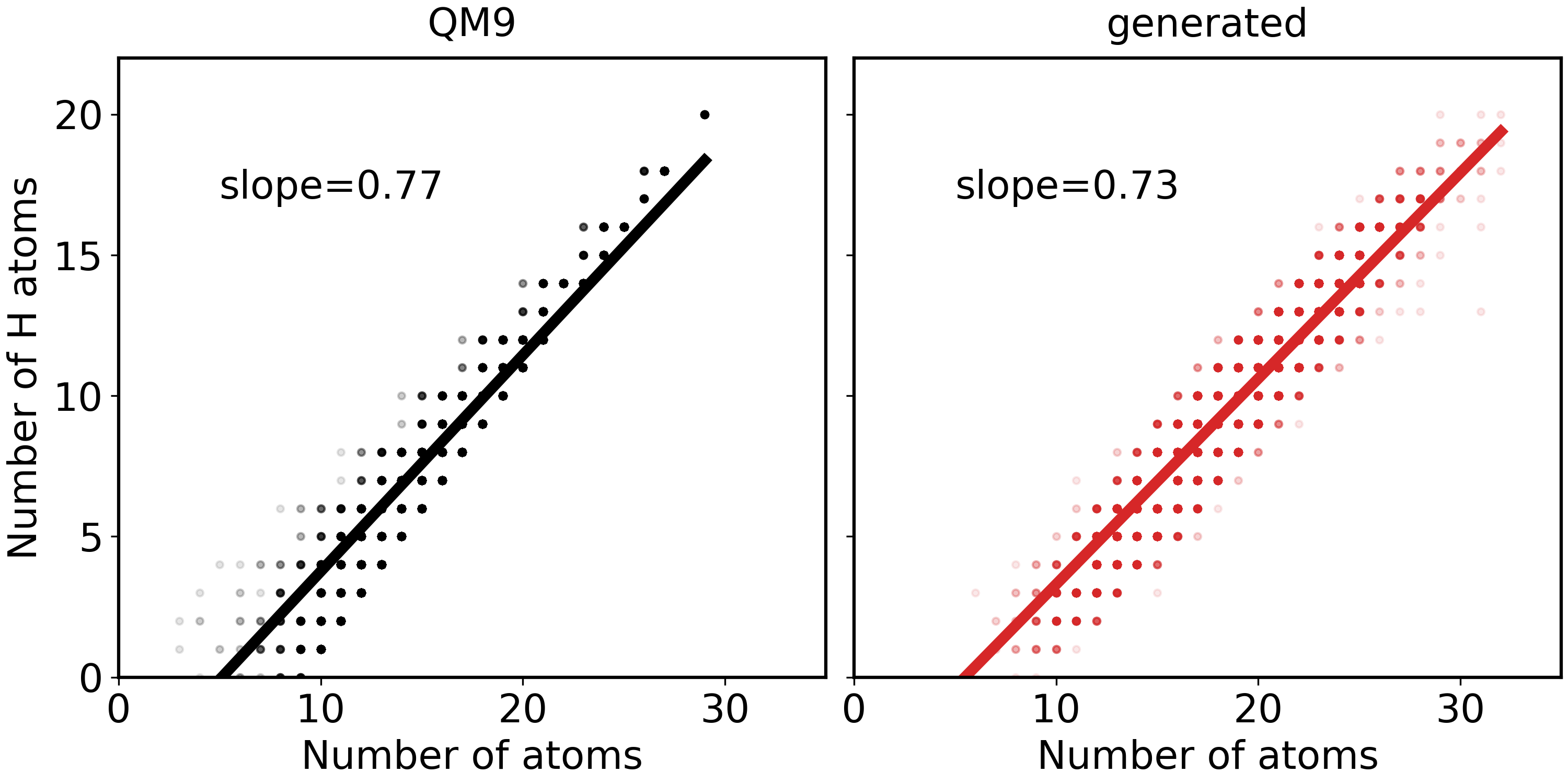}
    \caption{Number of H atoms as a function of number of atoms for the QM9 straining data (left) and the G-SchNet generated molecules (right). The dots indicate the distribution and the line corresponds to a trend line fitted to this distribution.}
    \label{fig:qm9_nH_natoms}
\end{figure}

\begin{figure}
    \centering
    \includegraphics[width=0.4\linewidth]{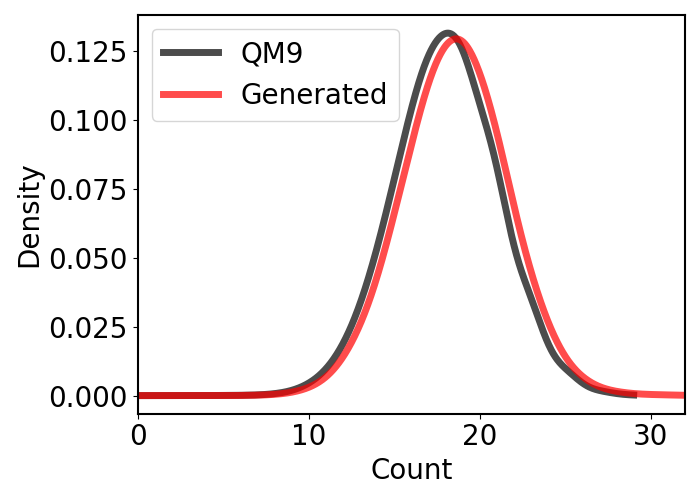}
    \includegraphics[width=0.4\linewidth]{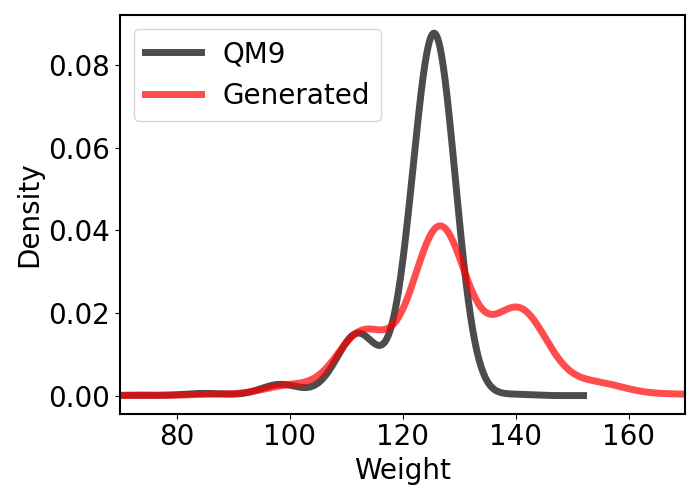}
        \includegraphics[width=0.4\linewidth]{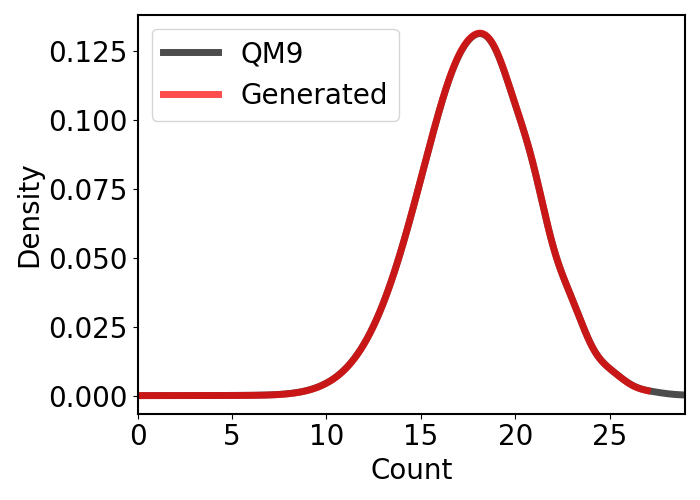}
    \includegraphics[width=0.4\linewidth]{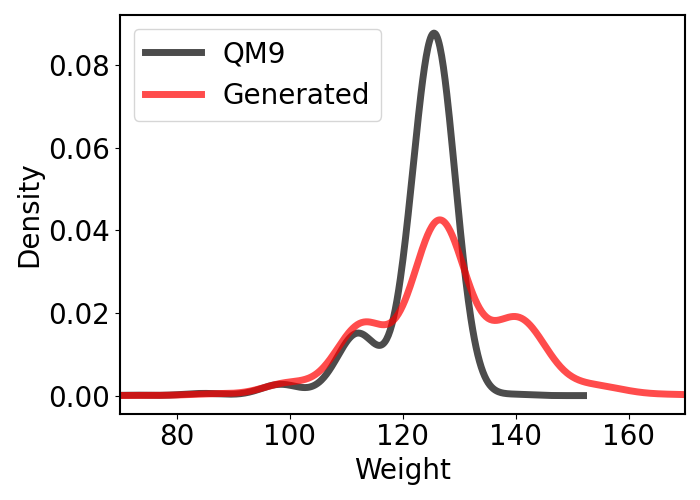}
    \caption{Distribution of total atom counts and molecular weights in QM9 training and generated molecules. Top panels: comparison of QM9 training data and original generated molecule database, Bottom panels: comparison of QM9 training data and the generated dataset after resampling according to the training set's atom count distribution.}
    \label{fig:qm9_natoms}
\end{figure}

\begin{figure}
    \centering
    \includegraphics[width=0.4\linewidth]{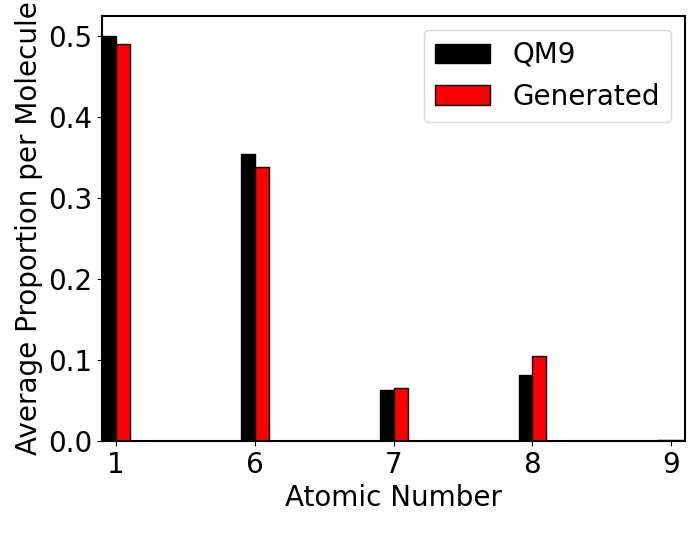}
    \includegraphics[width=0.4\linewidth]{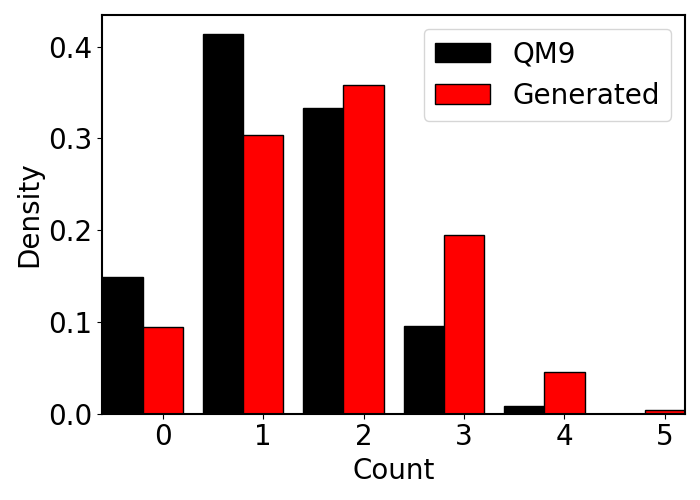}
    \caption{Left: Elemental composition of QM9 and generated molecular databases shown as average proportion of elements per molecule. Right: Frequency of occurrence of oxygen atoms in QM9 and generated molecules.
    In both cases, the generated database was resampled according to the training set's atom count distribution.}
    \label{fig:qm9_elements}
\end{figure}

\begin{figure}
    \centering
    \includegraphics[width=0.9\linewidth]{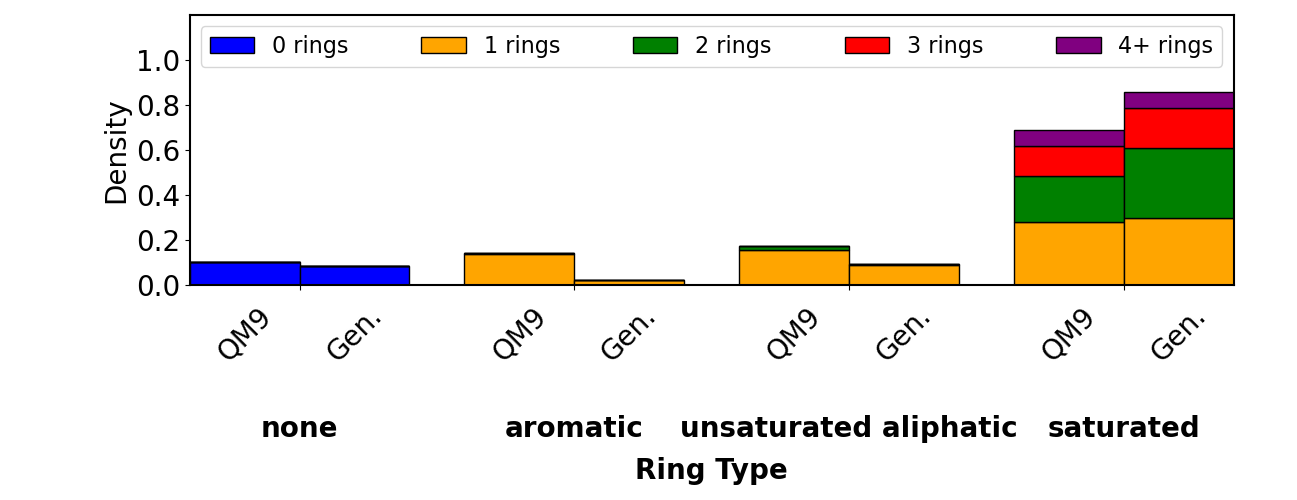}
    \caption{Proportion of QM9 training and generated molecules containing rings, separated into different ring types and counts. }
    \label{fig:rings_qm9}
\end{figure}

\clearpage
\subsection{Decision Trees}

\begin{figure}
    \centering
    \includegraphics[width=0.9\linewidth]{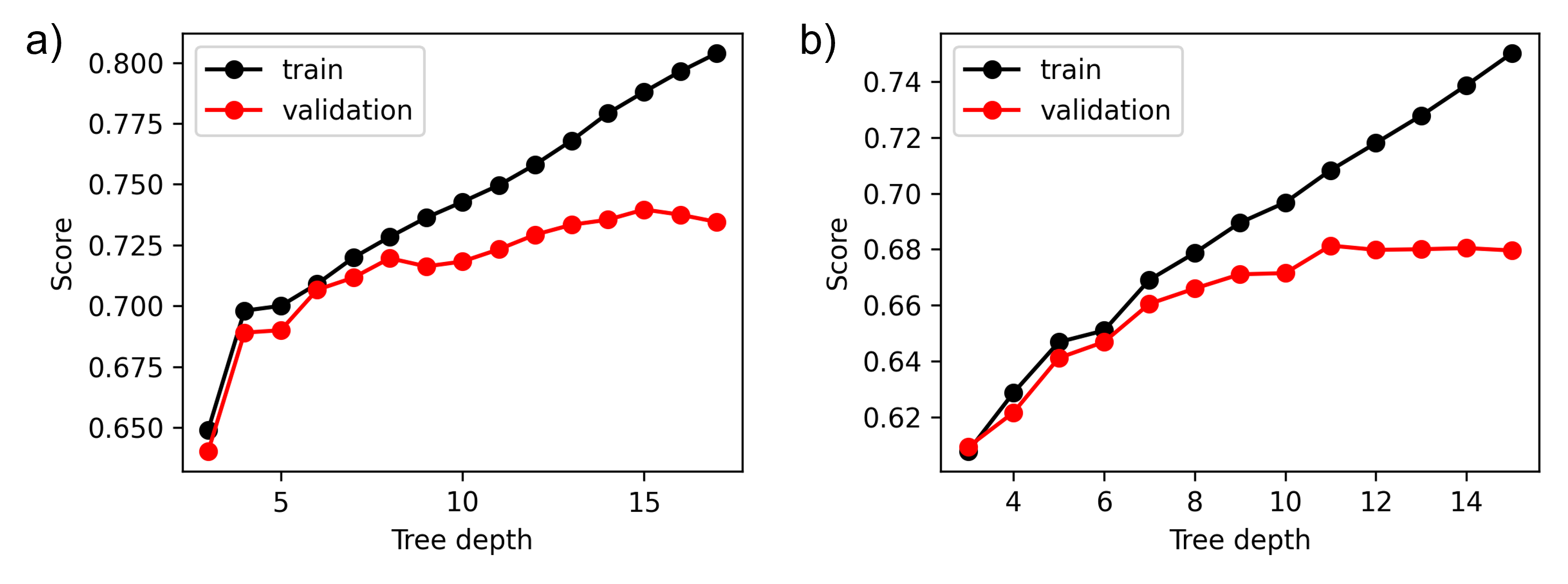}
    \caption{Optimising the tree depth parameter of a decision tree model to discriminate between QM9 and generated molecules. Features include a) Morgan fingerprints and atom counts b) Morgan fingerprints only. }
    \label{fig:qm9_tree}
\end{figure}

For training decision tree discriminators, we generated Morgan fingerprints with radius 0 and number of bits 1000. When also including atom counts in the descriptor, the optimal tree depth is 15 with validation score 0.740 (Figure \ref{fig:qm9_tree}). Without atom counts, the optimal tree depth is 11 with validation score 0.681. Thus including atom counts in the feature vector results in significantly more accurate predictions, and the former model is used in the analysis below.

\begin{figure}
    \centering
    \includegraphics[width=0.9\linewidth]{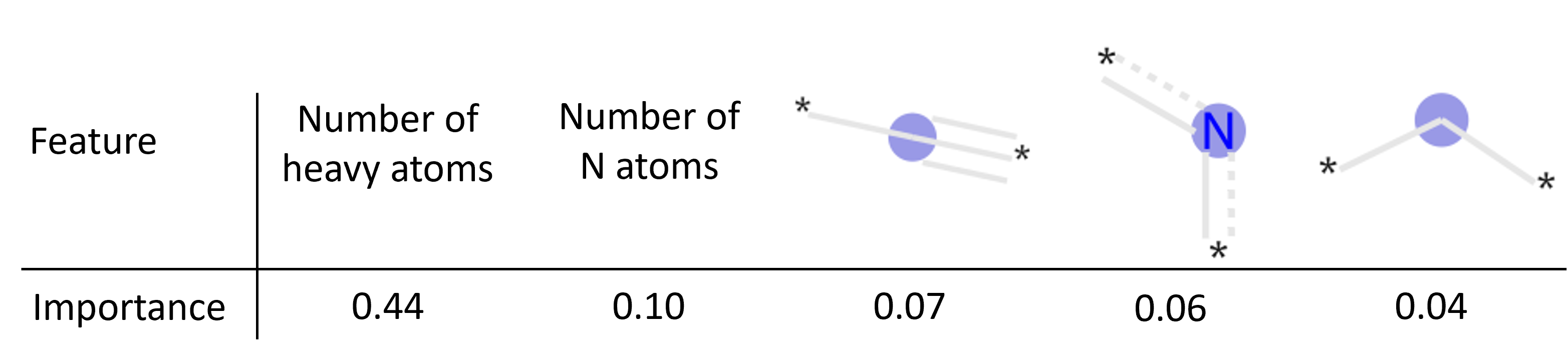}
    \caption{Top 5 most important features discriminating between QM9 and generated molecules using the optimal decision tree model. Purple circles mark the central atom in the Morgan fingerprint bits.}
    \label{fig:qm9_features}
\end{figure}

\clearpage

\section{Additional data for model trained on OE62 training data}

\subsection{Generated Molecule Analysis}

The OE62 training database contains 61489 molecules, out of which 53061 can be converted into valid RDKit molecules. 

\begin{figure}
    \centering
    \includegraphics[width=0.4\linewidth]{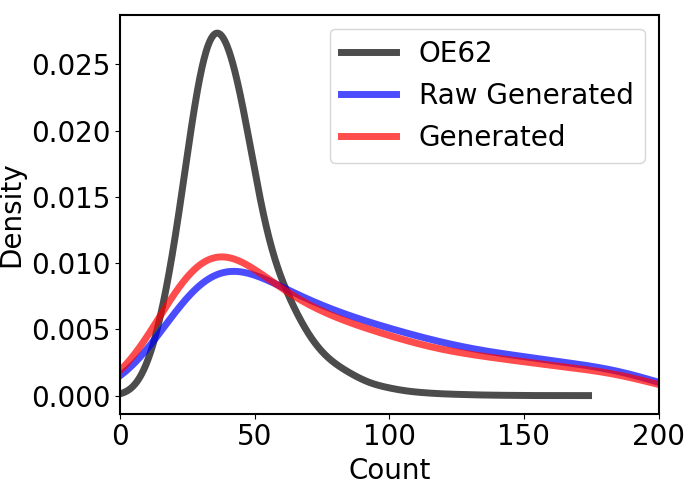}
    \includegraphics[width=0.4\linewidth]{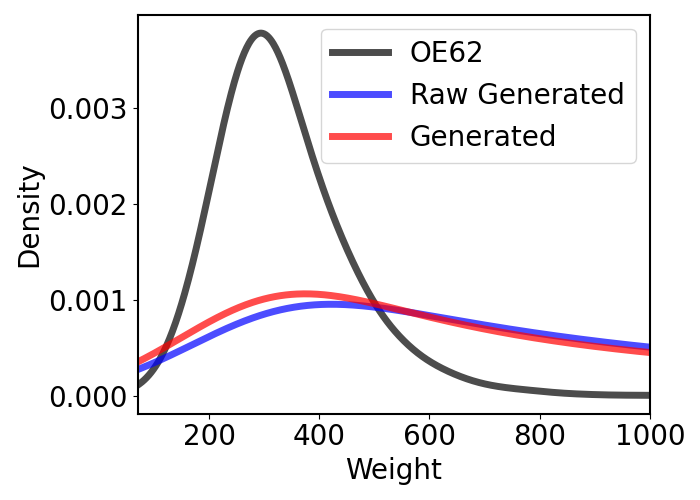}
    \includegraphics[width=0.4\linewidth]{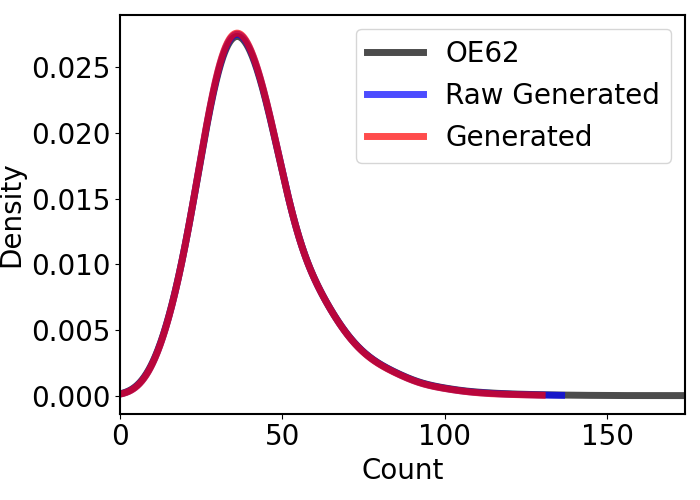}
    \includegraphics[width=0.4\linewidth]{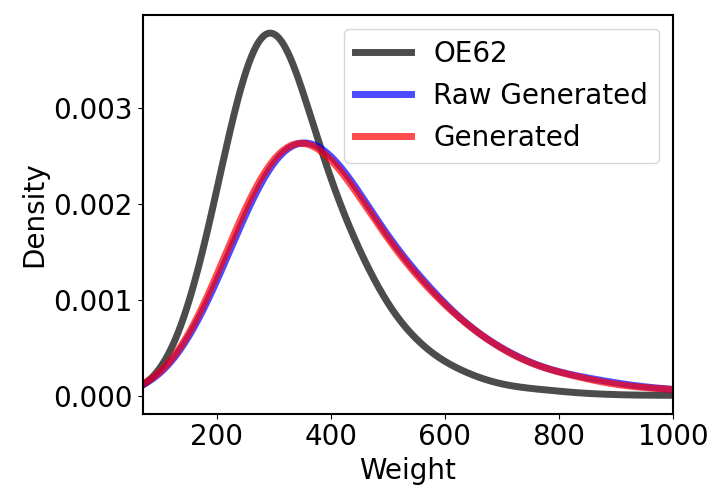}
    \caption{Distribution of total atom counts and molecular weights in OE62 training and generated (max. 200 atoms) molecules. Top panels: comparison of OE62 training data and original generated molecule database. The raw generated dataset refers to generated molecules without prior filtering of molecules according to connectivity and validity. Bottom panels: comparison of OE62 training data and the generated dataset after resampling according to the training set's atom count distribution.}
    \label{fig:oe62_natoms}
\end{figure}

\begin{figure}
    \centering
    \includegraphics[width=0.49\linewidth]{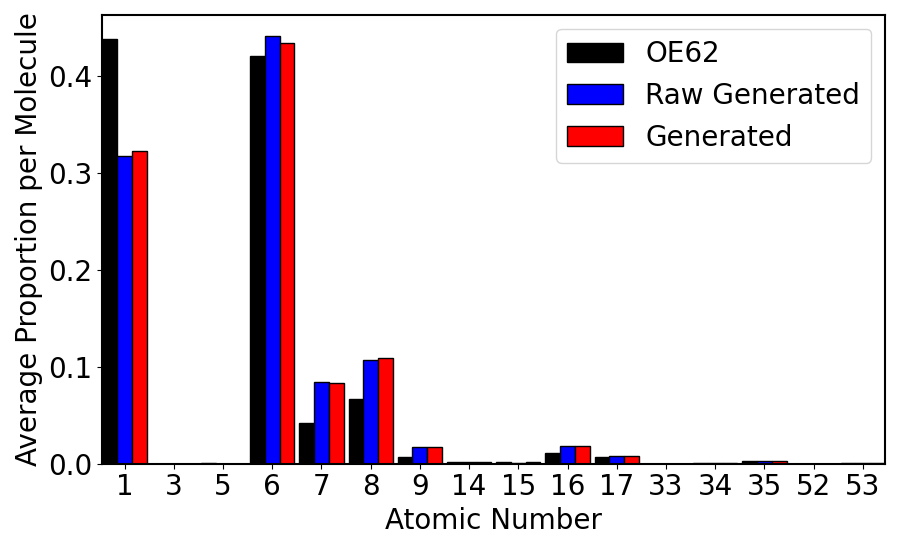}
    \includegraphics[width=0.49\linewidth]{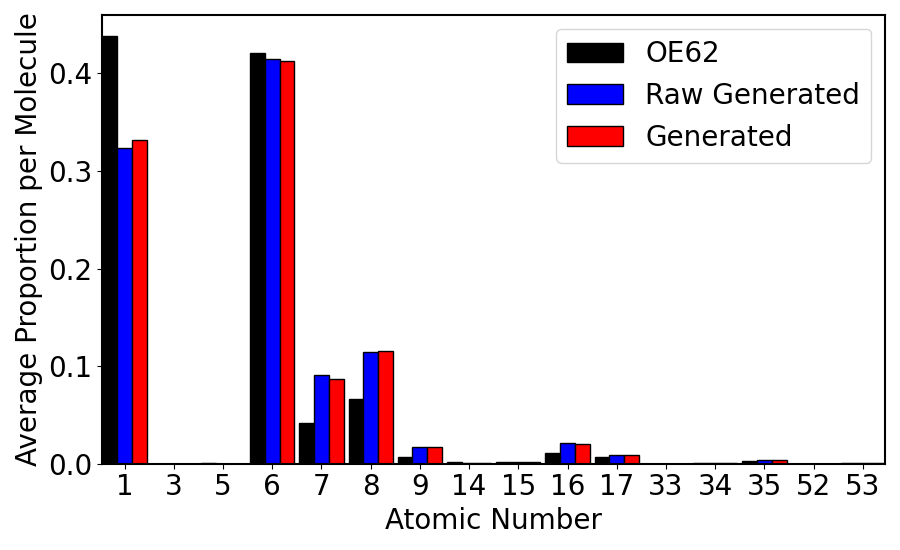}
    \includegraphics[width=0.48\linewidth]{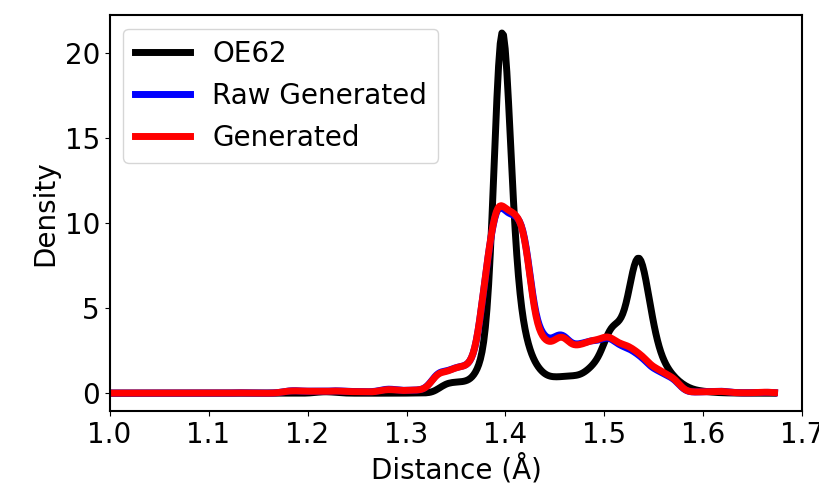}
    \caption{Top panels: Elemental composition of training data, raw generated and filtered generated structures. On the left, all generated molecules are used. On the right, the generated molecules were resampled according to the training set's atom count distribution.
    Bottom panel: Distribution of bonded C-C distances for the OE62 training set, the raw generated molecules, and the filtered generated molecules (both resampled).}
    \label{fig:oe62_elements}
\end{figure}

\begin{figure}
    \centering
    \includegraphics[width=0.9\linewidth]{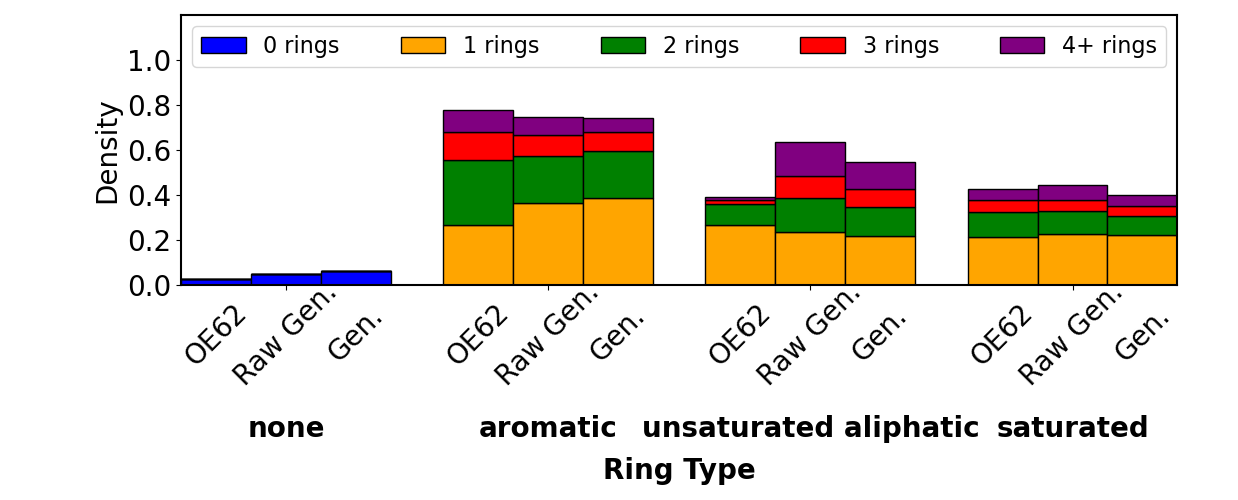}
    \includegraphics[width=0.9\linewidth]{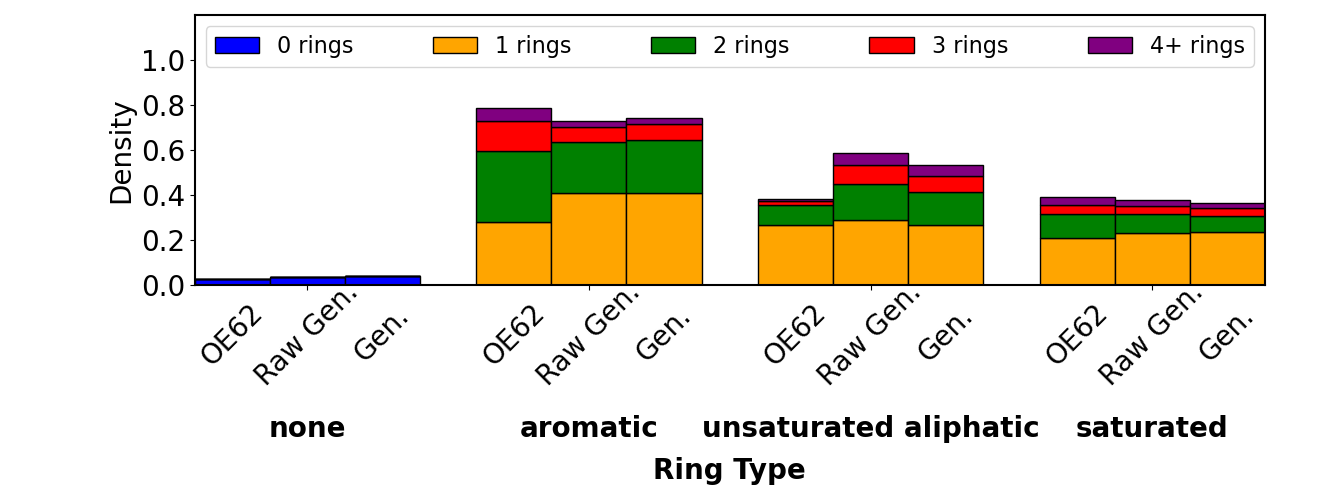}
    \caption{Proportion of molecules having different types and counts of rings in the OE62, raw generated and filtered generated databases. Top: all generated molecules, Bottom: the generated databases were resampled according to the training set's atom count distribution.}
    \label{fig:oe62_rings}
\end{figure}


\begin{figure}
    \centering
    \includegraphics[width=0.35\linewidth]{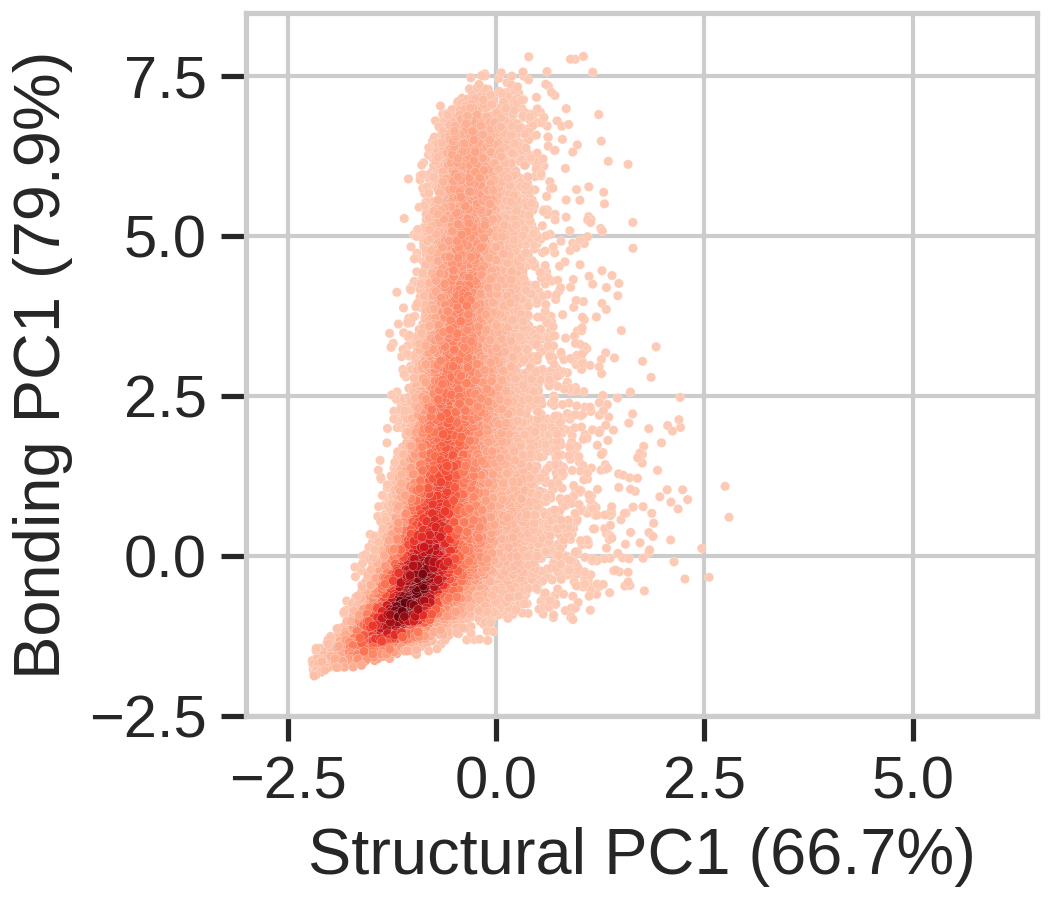}
    \caption{Latent chemical space covered by molecules generated by G-SchNet trained on OE62, filtered for validity but without any sampling with respect to the size distribution of the training database.}
    \label{fig:oe62_gen_pca_unsampled}
\end{figure}

\begin{figure}
    \centering
    \includegraphics[width=0.9\linewidth]{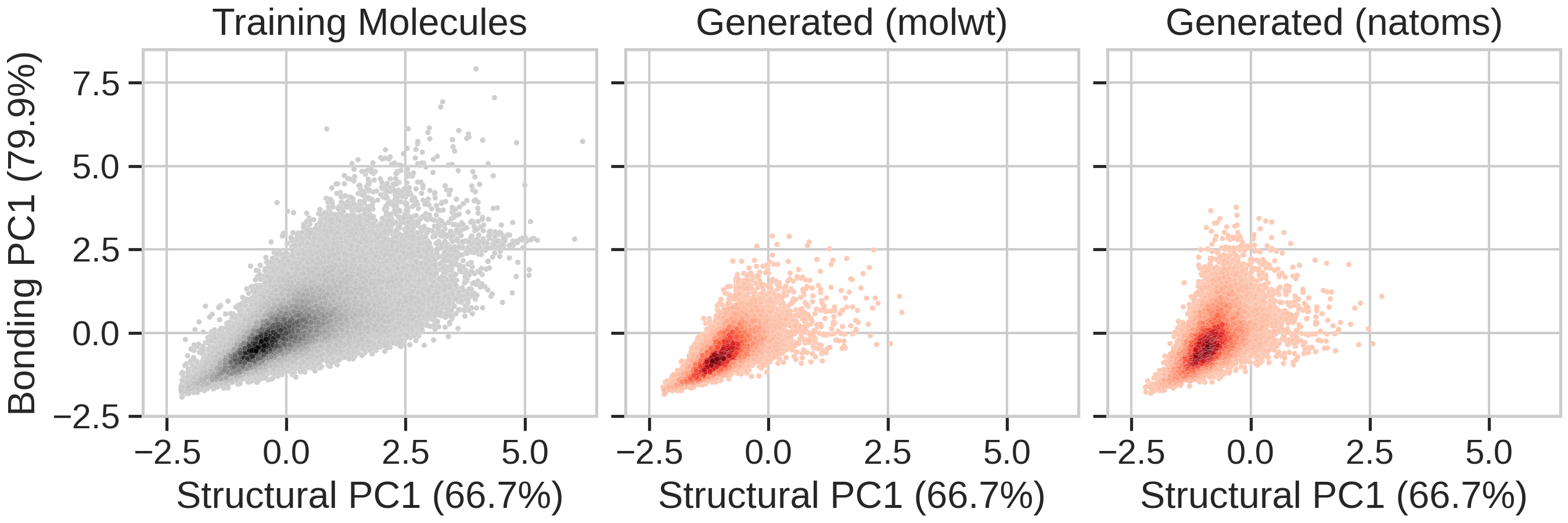}
    \caption{Latent chemical space covered by generated molecules after sampling with respect to molecular weight (molwt) and number of atoms (natoms) distributions in the OE62 training database.}
    \label{fig:oe62_pca_sampling_compare}
\end{figure}

\clearpage
\subsection{Improving G-SchNet}

\begin{table*}[]
    \centering
    \begin{tabular}{c|cccccc}
       & Target & Generated & Duplicate & Disconnected & Filtered & Valid RDKit\\
       \hline
        Original  & 60000 & 48051 & 0& 1556 & 25945 & 11233\\
        20 \AA\ cutoffs  &  60000& 39999 & 0 & 4750 & 19727 & 16578\\
        10 trajectories  & 60000& 20000 & 0 & 754 & 10529 & 8782\\
        1:3 loss ratio  &60000  & 19997 & 0 & 565 & 10625 & 8946\\
    \end{tabular}
    \caption{Number of unique and valid molecules generated with different G-SchNet settings, based on the OE62 training dataset.  }
    \label{tab:generated_stats}
\end{table*}


\begin{figure}
    \centering
    \includegraphics[width=0.46\linewidth]{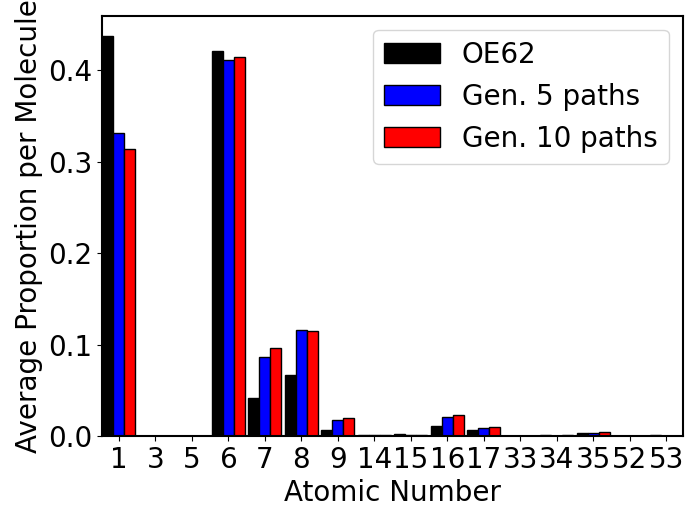}
    \includegraphics[width=0.48\linewidth]{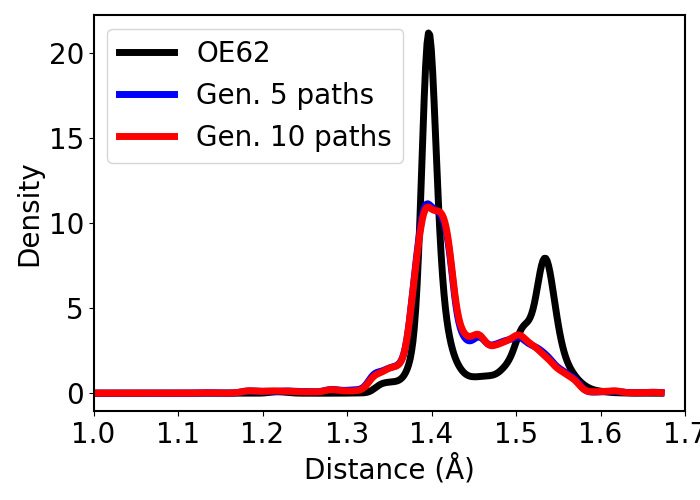}
    \caption{Elemental composition (left) and distribution of bonded C-C distances (right) for the OE62 training set and filtered generated molecules, with 5 or 10 random trajectories used for training G-SchNet, respectively. The generated databases were sampled according to the training set's atom count distribution.}
    \label{fig:oe62_elements}
\end{figure}

\begin{figure}
    \centering
     \includegraphics[width=0.46\linewidth]{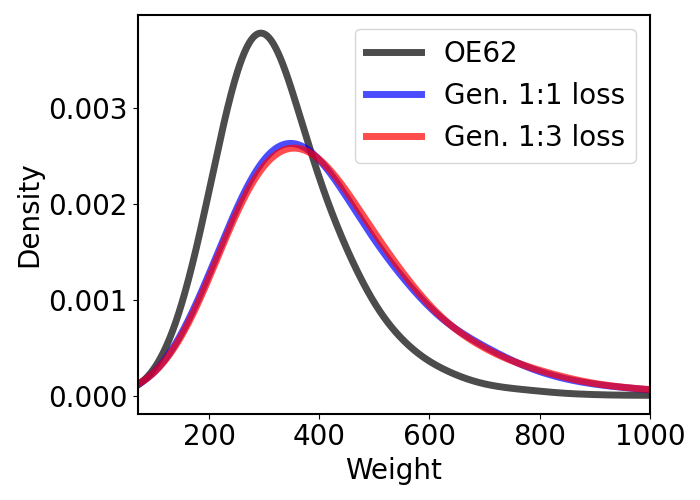}
    \includegraphics[width=0.46\linewidth]{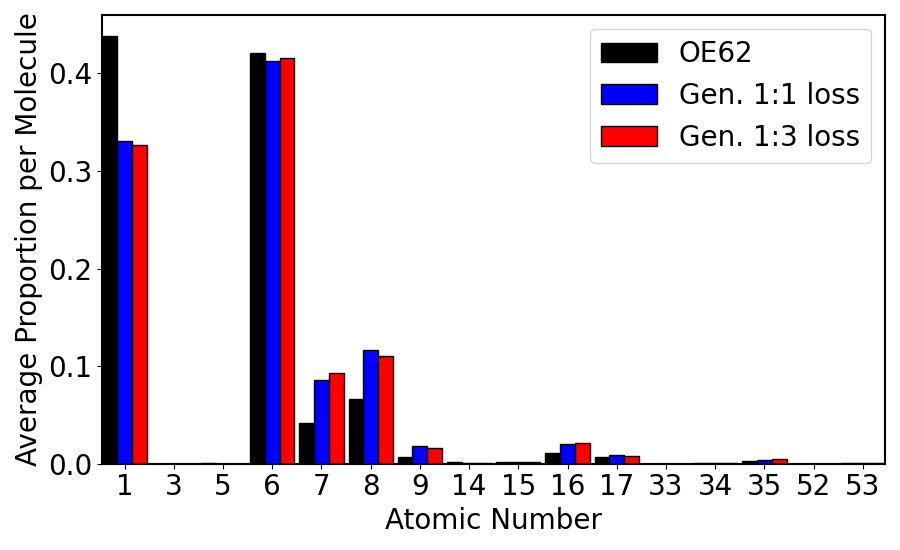}
    \includegraphics[width=0.48\linewidth]{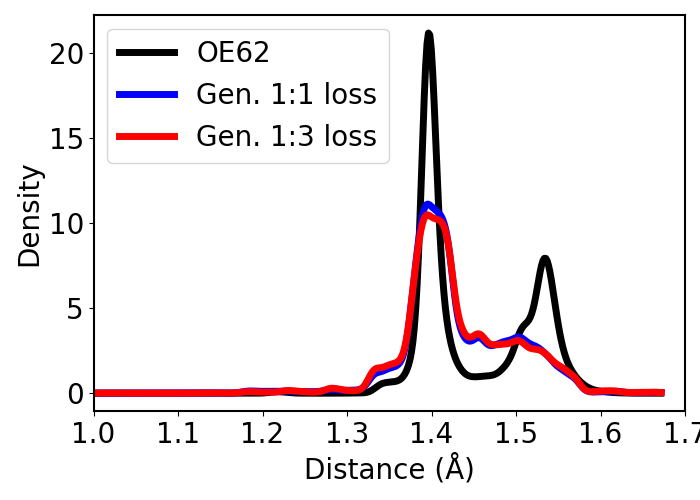}
    \includegraphics[width=0.48\linewidth]{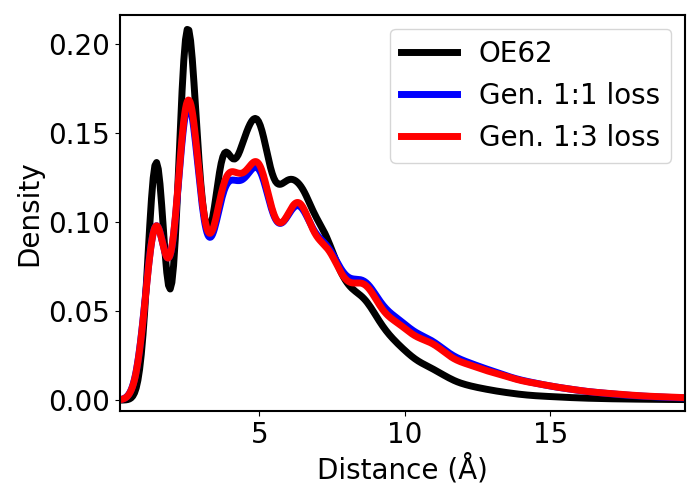}
        \includegraphics[width=0.48\linewidth]{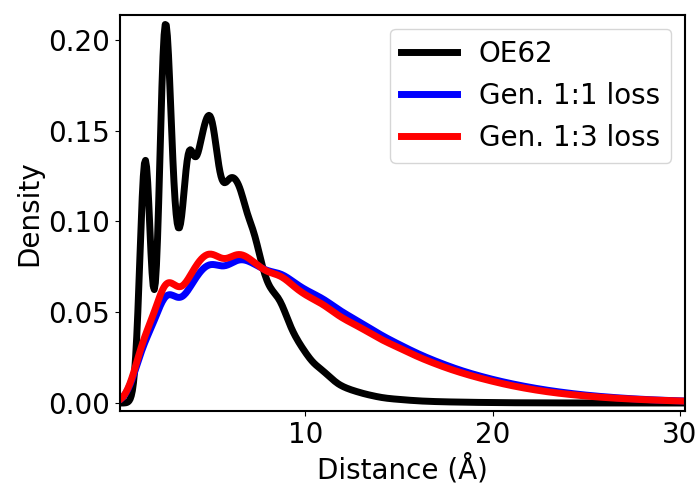}
    \caption{a) b) Elemental composition and c) distribution of bonded and d) all C-C distances for the OE62 training set and filtered generated molecules, with distance loss: element type loss ratios 1:1 or 1:3 used for training G-SchNet, respectively. The generated databases were sampled according to the training set's atom count distribution. e) all C-C distances, all generated molecules (no sampling).}
    \label{fig:oe62_elements}
\end{figure}

\begin{figure}
    \centering
    \includegraphics[width=0.5\linewidth]{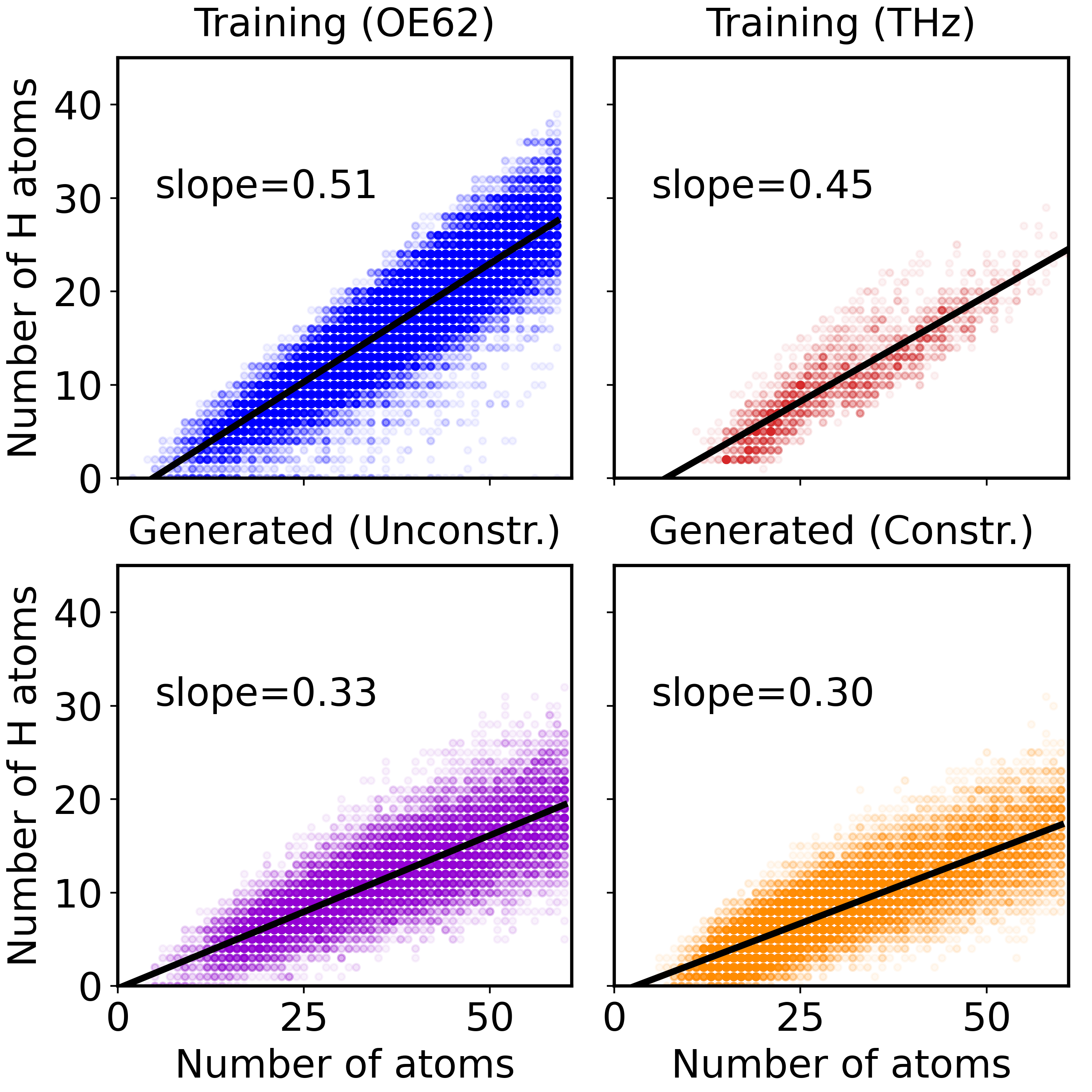}
    \caption{The effect of using a custom transform to enforce the thiolate group on the number of H atoms as a function of total number of atoms per molecule.}
    \label{fig:hatoms_constrained}
\end{figure}

\clearpage
\subsection{Decision Trees}

\begin{figure}
    \centering
    \includegraphics[width=0.9\linewidth]{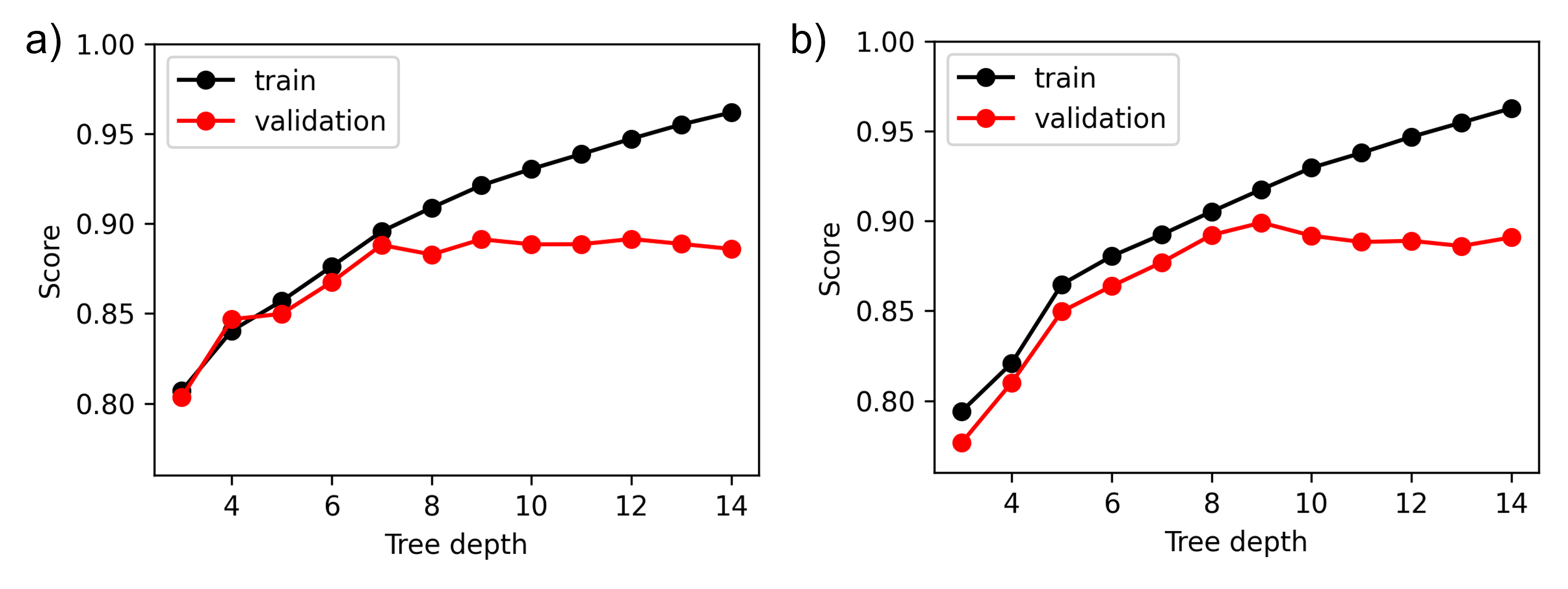}
    \caption{Optimising the tree depth parameter of a decision tree model to discriminate between OE62 and generated molecules. Features include a) Morgan fingerprints and atom counts b) Morgan fingerprints only. }
    \label{fig:oe62_features}
\end{figure}

When using Morgan fingerprints with radius 0 and number of bits 1000 combined with atom counts, the optimal tree depth is 13 with validation score 0.887. Without atom counts, the optimal tree depth is 9 with validation score 0.891. When using Morgan fingerprints with radius 1 and number of bits 2000, the best validation score is 0.891 for tree depth 14. Thus including atom counts in the feature vector or increasing the radius of Morgan fingerprints does not affect the score notably, and the simpler model without atom counts, fingerprints with radius 0 and number of bits 1000, and tree depth of 9 is used in further analysis. The five most important features are shown in \ref{fig:oe62_features}.

\begin{figure}
    \centering
    \includegraphics[width=0.9\linewidth]{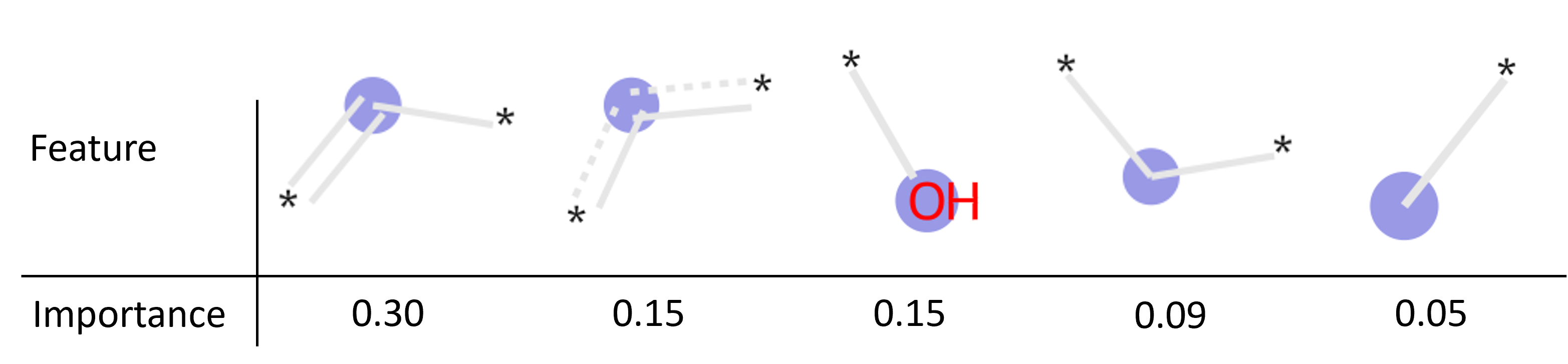}
    \caption{Top 5 most important features discriminating between OE62 and generated molecules using the optimal decision tree model. Purple circles mark the central atom in the Morgan fingerprint bits.}
    \label{fig:oe62_features}
\end{figure}
